\documentclass[aps,prd,10pt,twocolumn,superscriptaddress,nofootinbib,showkeys,showpacs,altaffilletter]{revtex4-1}

\usepackage{graphicx}
\usepackage{dcolumn}
\usepackage{amssymb}
\usepackage{amsmath}
\usepackage{amsfonts}
\usepackage{amsbsy}
\usepackage{color}
\usepackage{rotating}
\usepackage[english]{babel}
\usepackage{soul}
\usepackage{multirow}
\usepackage{mathtools}

\usepackage{hyperref}
\hypersetup{
    colorlinks=true,
    linkcolor=blue,
    filecolor=magenta,
    urlcolor=cyan,
    citecolor=cyan
}

\newcommand{\be}{\begin{equation}}
\newcommand{\ee}{\end{equation}}
\newcommand{\bea}{\begin{eqnarray}}
\newcommand{\eea}{\end{eqnarray}}
\newcommand{\beal}{\begin{aligned}}
\newcommand{\eeal}{\end{aligned}}
\newcommand{\bi}{\begin{itemize}}
\newcommand{\ei}{\end{itemize}}

\newcommand{\pa}{\partial}
\newcommand{\mt}{\mathcal}

\begin{document}

\title{Multi-component DHOST analysis in galaxy clusters}

\date{\today}

\author{Enrico Laudato}
\email{enrico.laudato@phd.usz.edu.pl}
\affiliation{Institute of Physics, University of Szczecin, Wielkopolska 15, 70-451 Szczecin, Poland}
\author{Vincenzo Salzano}
\email{vincenzo.salzano@usz.edu.pl}
\affiliation{Institute of Physics, University of Szczecin, Wielkopolska 15, 70-451 Szczecin, Poland}
\author{Keiichi Umetsu}
\email{keiichi@asiaa.sinica.edu.tw}
\affiliation{Academia Sinica Institute of Astronomy and Astrophysics (ASIAA), No. 1, Section 4, Roosevelt
Road, Taipei 10617, Taiwan}

\begin{abstract}
Extended Theories of Gravity with additional scalar degrees of freedom have recently acquired increasing interest due to the presence of a screening mechanism that allows suppressing at small scales (e.g., the Solar System scale) every modification restoring General Relativity. In this work, we consider a second-order Extended Theory of Gravity belonging to the family of Degenerate High Order Scalar Tensor theories (DHOST) characterized by a partial breaking of the Vainshtein screening mechanism. We study this model in two different scenarios as a description of dark energy only and as a description of both dark matter and dark energy. Such scenarios have been tested here by analysing a sample of 16 high-mass galaxy clusters targeted by the Cluster Lensing and Supernova survey with Hubble (CLASH) program using two complementary probes, namely X-ray and strong-and-weak gravitational lensing observations. In mass modelling, we adopt a multi-component approach including hot gas and galactic stellar contributions. For the majority of the clusters in our sample, results show mild Bayesian evidence in favour of the DHOST model as a description of dark energy over General Relativity. This model also appears to alleviate the discrepancy present in General Relativity between X-ray hydrostatic and lensing mass estimates. For the second scenario where gravity acts as both dark energy and dark matter due to the partial breaking of the Vainshtein screening mechanism at cluster scales, the model is statistically disfavoured compared  to General Relativity.
\end{abstract}


\maketitle

\section{Introduction}

At the present stage, we have no clue about what dark matter (DM) and dark energy (DE) should be. We do have a consensus model, the $\Lambda$ Cold Dark Matter ($\Lambda$CDM) paradigm, which is thus far the best effective model to explain the majority of data collected at both cosmological and astrophysical scales \cite{Akrami:2018vks,eBOSS:2020yzd,DES:2021wwk,DES:2021esc,vanUitert:2017ieu,Joudaki:2017zdt}. However, it has been recognized that the $\Lambda$CDM model is plagued by a number of theoretical problems \cite{Bull:2015stt} and observational conundrums, one of the most debated nowadays being the well-known $H_0$ tension \cite{DiValentino:2021izs,Freedman:2021ahq,Schoneberg:2021qvd,Efstathiou:2021ocp}.  

One of the main pillars of the $\Lambda$CDM model is the assumption that General Relativity (GR) is the ultimate theory of gravity. This is by far a ``huge'' inference itself, if we consider that GR has been well-tested only in a narrow scale range \cite{Will:2014kxa} of about four order of magnitudes. 
Ideally, applying GR on a range of scales of (at least) sixty orders of magnitude would be a proper (or less problematic) way to do definitive tests in cosmological context. 

The main problem of the $\Lambda$CDM model is that once GR is assumed, it is undoubtedly clear that something is missing: at least $95\%$ of matter and energy in the Universe should be present in certain ``dark'' components having totally different properties: DE should account for $\sim 68\%$ of the energy budget in the present Universe and cause its accelerated expansion \cite{SupernovaSearchTeam:1998fmf,SupernovaCosmologyProject:1998vns}; DM should account for $\sim 27 \%$ of the energy budget in the present Universe and be responsible for the extra gravitational attraction in astrophysical gravitational structures observed at galactic scales and beyond.
Here we only point out that none of the DM candidates (whose list is quite long \cite{Bertone:2004pz, Profumo:2019ujg}) has been detected so far, and we shall mainly focus on DE in this paper.

In the $\Lambda$CDM model, the most straightforward way to explain DE is to assume the cosmological constant $\Lambda$, which should be a measure of the energy density of the vacuum, its state of lowest energy \cite{Carroll:2000fy,Martin:2012bt}. The conditional mode ``should'' is obligatory because if $\Lambda$ was really the vacuum energy, then its value calculated from particle physics and quantum field theory does not agree with the observed one by $\sim122$ orders of magnitude. Despite this and other theoretical problems (for a more exhaustive list, see \cite{Bull:2015stt}), the cosmological constant is nowadays considered the best candidate for DE, for its intrinsic simplicity from both theoretical and statistical points of view, when a comparison with observational data is performed.

However, doubts have been cast on the $\Lambda$CDM model \cite{Bull:2015stt}, and cosmologists are looking for solutions. The most conservative alternatives still rely on GR: they save its geometrical part, but modify the energy-matter contribution, by adding phenomenologically one or more scalar fields with some given properties, and whose existence, however, has to be explained somehow. An absolutely non-exhaustive list can be found here \cite{Peebles:2002gy,Copeland:2006wr,Li:2011sd,Mortonson:2013zfa,Amendola:2016saw,Huterer:2017buf}. The less conservative approaches, instead, consider GR ``only'' as a special limit of a more general theory of gravity, so that now both geometry and matter are susceptible to modifications. Such theories are interchangeably called ``modified'', ``alternative'', or ``extended'' theories of gravity (ETGs) \cite{Clifton:2011jh,Ishak:2018his}. 

Attempting to put in order the ETGs zoo is quite challenging as tentatively sketched in \cite{Koyama:2015vza,Ishak:2018his}, because there are numerous ways in which extensions of GR can be built. However, this does not mean that it is straightforward to find a successful path. In fact, even once a theory passes all theoretical consistency tests (e.g., avoiding ghosts and instabilities \cite{DeFelice:2006pg}), it must satisfy observational tests: in particular, ETGs have to reduce to GR on scales where GR is found to be valid, as in the Solar System \cite{Will:2014kxa,Bertotti:2003rm,Shapiro:2004zz}. 
The ``safe exit'' mechanisms that allow ETGs to play the role of DE at cosmological scales and to reduce to GR in the correct limit are generally referred to as screening mechanisms. Once again, there are many different ways in which such screening mechanisms can be realized; see \cite{Joyce:2014kja,Burrage:2017qrf,Khoury:2003rn,Brax:2010kv,Brax:2010tj,Hinterbichler:2010es,Davis:2011pj,Brax:2012jr,Vainshtein:1972sx} for a review. In addition to that, the recent multi-messenger detection of  gravitational waves and the full electromagnetic spectrum counterpart emissions from a binary neutron star inspiral event GW170817 \cite{LIGOScientific:2017vwq,LIGOScientific:2017ync}, has imposed stringent limits on possible changes or modifications of gravity \cite{Creminelli:2017sry,Sakstein:2017xjx,Ezquiaga:2017ekz,Baker:2017hug,Baker:2020apq}


Among all the modified gravity theories that have passed all (or most) of such constraints, we will focus on specific family of theories with extra fields. Within this group, we have Horndeski models \cite{Horndeski:1974wa}, the most general single field scalar--tensor theory with second-order derivative equations of motion (which make them free of instabilities and other potentially pathological properties). These models can be extended to have higher-order derivative equations of motion, but with the degree of freedom which has only second-order field equations, and free of instabilities \cite{Deffayet:2011gz}. For such a reason, they are dubbed ``healthy'' extensions, and are called Beyond Horndeski theories \cite{Gleyzes:2014qga}. More recently, these same Beyond Horndeski theories have revealed to be a subset of a larger class of equally ``healthy'' higher-order theories. Such larger family is nowadays defined as Degenerate Higher-Order Scalar-Tensor (DHOST) theories \cite{BenAchour:2016fzp,Langlois:2018dxi}. From now on, when speaking about ETGs in our work, we will always refer to DHOST, in the sense of the largest possible class of theories with effective second-order field equations.

For theories with second (or higher) derivatives for the scalar field in the Lagrangian (as DHOST models), the screening mechanism that allows restoring the standard gravitational scenario at small scales is called the Vainshtein mechanism \cite{Vainshtein:1972sx} and is controlled by the kinetic contribution of the field, in particular when the second order derivatives
dominate. The most interesting point for our work, however, it that as pointed out in \cite{Kobayashi:2014ida}, for Beyond Horndeski and DHOST theories the Vainshtein screening can be partially broken. This implies that GR is restored only outside a possible overdensity (i.e. a galaxy, or a cluster of galaxies) while inside such structures, their dynamics and kinematics, which are tracers of the gravitational field, are not only influenced by the mass distribution (as it happens in GR), but also by new terms connected to such screening breaking.

Recently, we have begun to analyse a DHOST model characterized by a partially broken screening mechanism as developed in \cite{Koyama:2015oma,Sakstein:2016ggl}. To this end, we used in \cite{Salzano:2016udu,Salzano:2017qac} a sample of high-mass galaxy clusters with $M\sim 10^{15}M_\odot$ targeted by the Cluster Lensing and Supernova survey with Hubble (CLASH) program \cite{Postman:2011hg}. For these clusters, two independent sets of observational data are available to study their mass distributions. First, deep lensing observations with the \textit{Hubble Space Telescope} (\textit{HST}) were obtained for all 25 CLASH clusters \cite{Postman:2011hg}.  For a large subset of the CLASH sample, ground-based wide-field weak-lensing data are also available \cite{Umetsu2014,Merten2015}. Second, all CLASH clusters have observations with the \textit{Chandra} X-ray Observatory \cite{Donahue:2014qda}.

In this study, we aim to improve upon previous studies in two ways. First, we consider a DHOST model that encompasses the one analysed in \cite{Salzano:2016udu,Salzano:2017qac} but is more general in its formulation. Second, we model different matter components of CLASH clusters using observational results retrieved from the literature. Specifically, we account for the brightest cluster galaxy (BCG) and other galaxy components as well as the DM and hot intra-cluster gas.

Moreover, we will push our analysis toward a test of a ``new'' conjecture. In fact, DHOST theories, as well as most ETGs, are introduced as an alternative to GR to explain DE, but not DM. Here we will not only test this possibility but also attempt to answer the following question: what if DE and DM, through the breaking of the Vainshtein screening mechanism at astrophysical scales, could be unified in a single theoretical background? 

The paper is organized as follow: in Sec.~\ref{sec: Model} we introduce a theoretical model that represents the basis for our analysis and the theoretical background of the chosen set of observations; in Sec.~\ref{sec: Data} we present the data sets we consider and describe how we model the clusters; in Sec.~\ref{sec: Results} we present the results of our analysis and discuss their implications for DHOST theories; and, finally, in Sec.~\ref{sec: Conclusions} we draw our conclusions.

\section{Model}
\label{sec: Model}

Extending the analysis started in \cite{Salzano:2016udu,Salzano:2017qac}, we have decided to analyse the model described in \cite{Crisostomi:2016czh,Crisostomi:2017lbg,Dima:2017pwp}, which generalizes the model considered in the above works. We do have a DHOST theory in which the Vainshtein mechanism is turned on and partially broken within a gravitational system, and this can be readily seen by the expressions of the gravitational $(\Phi)$ and metric $(\Psi)$ potentials, derived in the weak field limit and assuming static spherical symmetry of the gravitational system:
\begin{eqnarray}
\label{eqn: model}
\frac{d\Phi}{dr} &=& \frac{G_NM(r)}{r^2} + \Xi_1 G_N M''(r)\, , \\
\frac{d\Psi}{dr} &=& \frac{G_NM(r)}{r^2} + \Xi_2 \frac{G_N M'(r)}{r} + \Xi_3 G_N M''(r)\, ,
\end{eqnarray}
where: $M(r)$ is the mass within the radius $r$, and its derivatives with respect to the distance are:
\begin{eqnarray}\label{eq:mass_derivatives}
M(r) &=& \int^{r}_{0} 4\pi r'^{2} \rho(r') dr'\, \nonumber \\
M'(r) &=& 4\pi r^2 \rho(r)\, \\
M''(r) &=& 8\pi r \rho(r) + 4\pi r^2 \frac{d \rho}{d r} \nonumber \,,
\end{eqnarray}
with $\rho$ the mass density of the system; $\Xi_1$, $\Xi_2$ and $\Xi_3$ (using the notation of \cite{Cardone:2020rmy}) are three dimensionless parameters which fully characterize the model and its deviation from GR, that can be recovered when $\Xi_{1,2,3} \rightarrow 0$; and $G_N$ is the measured effective gravitational constant, which might be different from the bare gravitational constant $G$ defined from $M_{PL}^{2} = \left(8 \pi G \right)^{-1}$. The fractional difference between $G$ and $G_N$ is generally expressed as a parameter as \cite{Dima:2017pwp}
\begin{equation}\label{eq:gamma0}
\gamma_0 = \left(8 \pi M_{PL}^{2} G_N \right)^{1}-1\, .
\end{equation}
Using the EFT approach \cite{Frusciante:2019xia,Piazza:2013coa,Bloomfield:2012ff}, the $\Xi$ parameters can be related to a more fundamental set of functions, initially used to describe Horndeski's theory \cite{Bellini:2014fua} and later extended to take into account also beyond Horndeski and DHOST theories \cite{Dima:2017pwp,Langlois:2017mxy,Langlois:2017dyl}. The most general expressions are \cite{Dima:2017pwp}:
\begin{align}
\label{eqn: xi param_0}
\Xi_1 &= -\frac{1}{2}\frac{(\alpha_H+ c_{T}^{2}\beta_1)^2}{c_{T}^{2} \left(1+\alpha_V-4\beta_1\right)-\alpha_H-1}\,,  \\
\Xi_2 &= -\frac{\splitfrac{\alpha_H\left(\alpha_H - \alpha_V +2 \left(1+c_{T}^{2}\right)\beta_1\right)+}
{\beta_1\left(c_{T}^{2}-1\right)\left(1+c_{T}^{2}\beta_{1}\right)}}{c_{T}^{2}\left(1+\alpha_V-4\beta_1\right)-\alpha_H-1}\, , \\
\Xi_3 &= -\frac{(\alpha_H+c_{T}^{2}\beta_1)}{c_{T}^{2}\left(1+\alpha_V-4\beta_1\right)-\alpha_H-1}\, ,\\
\gamma_0 &= \alpha_V - 3 \beta_1\, , 
\end{align}
where: $c_{T}^{2}$ is tensor speed excess and expresses the fractional difference between the speed of gravitons (GW) and of light; $\alpha_H$ measures the kinetic
mixing between matter and the scalar field introduced in Horndeski \cite{Horndeski:1974wa}, beyond Horndeski \cite{Zumalacarregui:2013pma,Crisostomi:2016tcp,Kobayashi:2019hrl}, and DHOST \cite{Dima:2017pwp} theories and characterizes the departure of the latter ones from Horndeski; $\alpha_V$ can be related to nonlinear dark energy perturbations \cite{Cusin:2017mzw}; and $\beta_1$ parameterizes the presence of higher-order
operators in the lagrangian \cite{Dima:2017pwp}. 

Such expressions can be heavily simplified once we take into account the strong constraints on alternative gravity theories derived from the multi-messenger observation of GW170817 \cite{Creminelli:2017sry,Sakstein:2017xjx,Ezquiaga:2017ekz,Baker:2017hug,Baker:2020apq}. The limit on the GW speed \cite{LIGOScientific:2017ync} alone is able to set $c_T=1$ and $\alpha_V = -\alpha_H$\footnote{Another possible DHOST scenario, in which is also required that gravitons do not decay into DE, is highlighted in \cite{Crisostomi:2019yfo}.}, which bring to the relations that we will use in our analysis, i.e.
\begin{eqnarray}
\label{eqn: xi param}
\Xi_1 &=& -\frac{1}{2}\frac{(\alpha_H+\beta_1)^2}{\alpha_H+2\beta_1}\, ,\\
\Xi_2 &=& \alpha_H\, ,\\
\Xi_3 &=& -\frac{\beta_1}{2}\frac{(\alpha_H+\beta_1)}{\alpha_H+2\beta_1}\, , \\
\gamma_0 &=& - \alpha_H - 3 \beta_1\, .
\end{eqnarray}
Something which might be hidden and it is worth to stress explicitly is that in principle we could have worked with the following set of free parameters, $\{\alpha_H, \beta_1\, \gamma_0\}$, or equivalently $\{\alpha_H, \beta_1, G\}$, because from Eq.~(\ref{eq:gamma0}) we have that:
\begin{equation}\label{eq:GN_G}
G_N = \frac{G}{1+\gamma_0}\, .
\end{equation}
Instead we will assume that $G_N$ is fixed to its measured value, so that $\gamma_0$ is fully determined by $\alpha_H$ and $\beta_1$ only, while $G$ is to be eventually determined from them.

\subsection{Gravitational lensing}

In a typical configuration of gravitational lensing \cite{Bartelmann:1999yn,Umetsu:2020wlf}, we consider a source positioned at an angular diameter distance from the observer $D_{s}$ and a lens at a distance $D_{l}$, with the distance between the lens and the source generally denoted as $D_{ls}$. When working with astrophysical lenses such as clusters of galaxies, we generally assume the weak-field limit and that the peculiar velocity of the lens is small. Moreover, given the difference in scales between the observer--lens and lens--source distances and the physical dimensions of the lens, we assume that this system can be approximately considered as a two-dimensional one (``thin-lens'' approximation). The main effect of the lens in such a regime is to deflect light rays from the source by a certain angle $\hat{\vec{\alpha}}$, which, in GR, can be defined as
\be
\hat{\alpha} = \frac{2}{c^2}\int^{+\infty}_{-\infty}dz\vec\nabla_\perp\Phi\,
\ee
where $\vec\nabla$ is the two-dimensional gradient operator orthogonal to the propagation of light and $z$ is the coordinate along the propagation direction.
The deflection angle $\hat\alpha$ can then be expressed in terms of the effective lensing potential, $\Phi_\mathrm{lens}$, i.e. the line-of-sight projection onto the lens plane of the full three-dimensional potential of the cluster \be
\label{eqn: lens}
\Phi_\mathrm{lens}(R) = \frac{2}{c^2}\frac{D_{ls}}{D_lD_s}\int^{+\infty}_{-\infty}\Phi(R,z)dz \, ,
\ee
where $R$ is the two-dimensional projected radius in the lens plane.

Applying the Laplacian operator on such a potential leads to the lensing convergence, $\kappa$, defined as
\be
\kappa(R) = \frac{1}{c^2}\frac{D_{ls}D_l}{D_s}\int^{+\infty}_{-\infty}dz\Delta_r\Phi(R,z)
\ee
where $r = \sqrt{R^2+z^2}$ is the three-dimensional radius and $\Delta_r = \frac{2}{r}\frac{\pa}{\pa r} + \frac{\pa^2}{\pa^2 r}$ is the radial Laplacian in spherical coordinates (other components are washed out by assuming spherical symmetry). Using the Poisson equation
\be
\Delta_r\Phi = 4\pi G_N\rho(r),
\ee
we can relate the convergence $\kappa$ to the mass density distribution $\rho$ of the lens system (with respect to the mean matter density of the universe) as
\be
\label{eqn: kappa}
\kappa(R) = \frac{4\pi G_N}{c^2}\frac{D_{ls}D_l}{D_s}\int^{+\infty}_{-\infty}dz\rho(R,z)\equiv\frac{\Sigma}{\Sigma_c},
\ee
where the two-dimensional surface density $\Sigma$ of the lens is defined as
\be
\Sigma = \int^{+\infty}_{-\infty}dz\rho(R,z)
\ee
and the critical surface mass density of gravitational lensing is
\be
\Sigma_c = \frac{4\pi G_N}{c^2}\frac{D_{ls}D_l}{D_s}.
\ee
Thus far, we have always assumed GR, i.e. $\Phi= \Psi$. However, in case of a generalised gravity theory, we need to modify the expression based on the fact that it can be $\Phi \neq \Psi$. In this more general case, the convergence is defined as
\be
\label{eqn: kappamod}
\kappa(R) = \frac{1}{c^2}\frac{D_{ls}D_l}{D_s}\int^{+\infty}_{-\infty}dz\Delta_r\biggl\{\frac{\Phi(R,z) + \Psi(R,z)}{2}\biggr\}\, .
\ee
It is important to stress that to determine the convergence $\kappa(R)$ it is crucial to specify the cosmological background, involved in the determination of the various angular diameter distances
\be
\label{eqn: distance}
D_A(z) = \frac{c}{1+z}\int_0^z \frac{dz'}{H(z)}\, ,
\ee
where $H(z)$ is the Hubble function. As we are not concerned with a cosmological analysis of the chosen DHOST model \cite{Crisostomi:2017pjs,Hiramatsu:2020fcd}, we simply assume that the background cosmology is well described by the \textit{Planck} baseline model $2.40$\footnote{\url{https://wiki.cosmos.esa.int/planck-legacy-archive/images/2/21/Baseline_params_table_2018_95pc_v2.pdf}} 
characterized by the Hubble constant $H_0 = 67.89$ and the matter density parameter $\Omega_m = 0.308$.

\subsection{X-ray hot gas}

In the approach based on X-ray observations, the main hypothesis is the assumption of spherical symmetry and, more importantly, the assumption that the system is in hydrostatic equilibrium. From the moments of the collisionless Boltzmann equation, we have the following equation:
\begin{equation}\label{eq:boltzmann_equation}
-\frac{{\mathrm{d}}\Phi(r)}{{\mathrm{d}}r} = \frac{k T_\mathrm{gas}(r)}{\mu m_{p} r}\left[\frac{{\mathrm{d}}\ln\rho_\mathrm{gas}(r)}{{\mathrm{d}}\ln r} +\frac{{\mathrm{d}}\ln T_\mathrm{gas}(r)}{{\mathrm{d}}\ln r}\right] \; ,
\end{equation}
from which, in GR, we derive:
\begin{eqnarray}\label{eq:mass_boltzmann_equation}
M_\mathrm{tot}(r) &=& M_\mathrm{gas}(r) + M_\mathrm{gal}(r) + M_\mathrm{BCG}(r) + M_\mathrm{DM}(r) \nonumber \\
&=& -\frac{k T_\mathrm{gas}(r)}{\mu m_{p} G_{N}} r \left[\frac{{\mathrm{d}}\ln\rho_\mathrm{gas}(r)}{{\mathrm{d}}\ln r}+\frac{{\mathrm{d}}\ln T_\mathrm{gas}(r)}{{\mathrm{d}}\ln r}\right] \, .
\end{eqnarray}
Thus, from the observed gas density and temperature profiles, it is possible to infer the total mass in the cluster, $M_\mathrm{tot}$. By measuring the gas mass from the same set of observations and combining with the baryonic mass contained in cluster galaxies, we can infer the DM distribution in the cluster, $M_\mathrm{DM}$. The total mass $M_\mathrm{tot}$ is generally considered to be thermal (or gravitational), but other non-thermal (or non-gravitational) phenomena might be at work, mostly in the innermost region of the clusters. Finally, what is critically important for this study is that the observationally derived total mass can be used to constrain the properties of the DHOST model using the following relation:
\begin{equation}\label{eq:total_mass_mod}
M^\mathrm{obs}_\mathrm{tot} = \frac{r^2}{G_{N}} \equiv M^\mathrm{theo}_\mathrm{tot} \frac{\mathrm{d}\Phi}{\mathrm{d}r}\; ,
\end{equation}
where the right-hand side is provided by Eq.~(\ref{eqn: model}).

\section{Data}
\label{sec: Data}

The data sets we use for our analysis are derived from the CLASH program \cite{Postman:2011hg}.
The sample of clusters targeted by the CLASH program is subdivided into two subsamples: (i) 20 X-ray-selected clusters having temperatures exceeding 5~keV and nearly concentric X-ray isophotes with a well-defined X-ray peak closely located to the BCG position; and (ii) five clusters selected by their exceptional lensing strength so as to magnify galaxies at high redshift. These high-magnification-selected systems often turn out to be dynamically disturbed, highly massive ongoing mergers \cite{Umetsu:2020wlf}. In contrast, numerical simulations in the $\Lambda$CDM framework suggest that the CLASH X-ray-selected subsample is prevalently composed of relaxed clusters ($\sim 70\%$) \cite{Meneghetti2014},  but it also contains a
non-negligible fraction ($\sim 30\%$) of unrelaxed clusters.

All 25 CLASH clusters have \textit{HST} strong- and weak-lensing data and resulting mass models in their central regions. Umetsu et~al. \cite{Umetsu:2015baa} reconstructed binned convergence profiles $\kappa(R)$ for 16 X-ray-selected and 4 lensing-selected CLASH clusters. The lensing analysis of \cite{Umetsu:2015baa} combines wide-field shear and magnification weak-lensing constraints primarily from the Subaru telescope \cite{Umetsu2014} and small-scale strong-and-weak lensing constraints from \textit{HST} \cite{Zitrin2015}.

In this work, we shall focus on a subsample of 13 X-ray-selected and 3 lensing-selected CLASH clusters from \cite{Umetsu:2015baa} (see Table~\ref{tab:results}).  For these CLASH clusters, wide-field weak-lensing measurements extending beyond the virial radius are available \cite{Umetsu2014} in addition to the \textit{HST} lensing data products of \cite{Zitrin2015}. Here we have discarded three X-ray-selected (A383, MACSJ1931, and RXJ1532) and one lensing-selected (MACSJ0717) CLASH clusters from the \cite{Umetsu:2015baa} sample, because the photometric data and/or mass estimates for their BCGs were not available, following the criteria we will describe in more detail in Sec.~\ref{sec:BCG}. According to \cite{Meneghetti2014}, about half of our sample clusters are expected to be unrelaxed, for which the assumption of hydrostatic equilibrium is not strictly satisfied.

For our X-ray analysis, we use the \textit{Chandra} X-ray data products (gas density and temperature in spherical bins for each cluster) obtained by \cite{Donahue:2014qda}. Our CLASH subsample spans a redshift range of $0.19 \le z \le 0.69$ with a median redshift of $0.352$. 
The resolution limit of mass reconstruction set by the \textit{HST} lensing data is $10$~arcseconds, which corresponds to $\approx 35h^{-1}$~kpc at the median redshift $z=0.352$ of the sample. Since this scale is larger than the typical harlf-light radii of the CLASH BCGs \cite{Tian:2020qjd}, CLASH lensing does not spatially resolve the BCGs. In \cite{Umetsu:2015baa}, it was found that the stacked lensing signal $\kappa(R)$ of the CLASH X-ray-selected subsample is best described by the Navarro--Frenk--White (NFW) model \cite{Navarro:1995iw} in a GR context (see also \cite{Umetsu:2016cun}).



As long as one considers GR or even the DHOST theory described above but including a DM component, the weight of components others than DM or gas will be of course quite negligible, see also Fig.~(\ref{fig:mass_comp}) for some examples. DM is the dominant component all over the cluster scales, ranging from $30\%$ to $60\%$ (of the total mass) at $5$~kpc to $5$~Mpc; while the hot gas  ranges from $\sim5\%$ to $\sim40\%$ in the same radial range. However, its statistical weight and influence is far from trivial: it is well known that X-ray hot gas is highly influenced by non-gravitational and local astrophysical phenomena mainly in the inner regions of the clusters ($<100$ kpc), and these effects may lead to biased estimates (up to $10-20 \%$) of the total mass of the cluster. By contrast, lensing observations (in this case, dominated by the DM distribution) are less prone to such astrophysical influences and can provide a direct mass and gravitational potential reconstruction of the system. These two characteristic are generally expressed by the fact that X-ray emissions from non-relativistic massive objects are sensitive to the Newtonian gravitational potential $\Phi$, while lensing, connected to the propagation of photons, is sensitive to the combination $\Phi+\Psi$. 

Yet, we must consider that in the innermost region of clusters, BCGs can give non-negligible contributions: from $\sim50 \%$ at $10-20$~kpc, where they are comparable to DM and dominant with respect to gas; to $\sim 10\%$ at $50-100$~kpc, where they are subdominant with respect to DM but still comparable to the hot gas. Thus, the inclusion of BCGs might have some consequences on the determination of the DHOST parameters and even in GR. In the latter case, we expect it to be fairly small, because the BCG contribution is dominant only in a limited radial range compared to DM and gas, which should be statistically dominant in the global mass profile. The non-BCG galaxy contribution is much more diffuse than BCGs and subdominant compared to other components, being $\sim1-2 \%$ in the radial range $20-300$~kpc; nevertheless, we include the non-BCG galaxy contribution in our analysis for the sake of completeness.

The ``power relations'' are totally reverted when we consider the DHOST as both DE and DM candidates. In such a scenario, we have no more a physical DM component, so that the hot X-ray emitting gas represents the most dominant contribution, being $>80\%$ at cluster-centric distances $>200$~kpc, while the BCGs dominate over or are comparable to the hot gas up to $100$~kpc.

The DHOST parameters must be carefully calibrated to provide a proper fit over the full range of scales. In fact, the DHOST contributions from the broken screening mechanism strongly depend on the mass density profiles. Thus, the effects of BCGs, which are dominant in the innermost region of clusters, are strongly accentuated by the DHOST parameters; and they might not agree with the effects of hot gas, which cover much larger distances, up to Mpc scales, and follow a different density profile. Taking into account that at $100$~kpc other galaxies also add a contribution, we understand that fitting DHOST to cluster mass profiles requires a careful modelling of each component and a detailed balance of each contribution for the goodness of the final statistical analysis. 

\subsection{Dark matter}

When a DM component is considered (in GR analysis and when the DHOST theory is assumed to be only a DE candidate), we assume the standard spherically symmetric NFW density profile \cite{Navarro:1995iw},
\be
\rho_\mathrm{NFW}(r) = \frac{\rho_s}{\frac{r}{r_s}\bigl(1 + \frac{r}{r_s}\bigr)^2}\, ,
\ee
where $\rho_s$ and $r_s$ are the characteristic NFW density and radius respectively. 
The characteristic NFW density $\rho_s$, is expressed as
\begin{equation}\label{eq:rhos500}
            \rho_s = \frac{\Delta}{3}\rho_c\, \frac{c_{\Delta}^3}{\ln(1+c_{\Delta})-\frac{c_{\Delta}}{1+c_{\Delta}}},
\end{equation}
where all quantities with subscript $\Delta$ are evaluated at the spherical radius $r_{\Delta}$, at which the mean interior density of the system is $\Delta$ times the critical density of the Universe at the same redshift of the lens. In our case, we consider the value $\Delta=500$. For cluster-scale haloes, $r_{500}$ is typically about half of the virial radius.
The concentration parameter $c_{\Delta}$ which appears in Eq.~(\ref{eq:rhos500}) is defined as
\begin{equation}\label{eq:c500}
c_{\Delta} = \frac{r_{\Delta}}{r_s},        
\end{equation}        
where $r_{\Delta}=r_{500}$ in our case. Finally, the free parameters for the NFW component will be $\{c_{500},r_{500}\}$.

\subsection{Hot gas}

Regarding the X-ray emitting hot intra-cluster gas, we follow \cite{Donahue:2014qda} and fit the gas mass profile using a double $\beta$-model \cite{1978A&A....70..677C}
\be
\label{eqn: beta model}
\beal
\rho_\mathrm{gas}(r) =& \rho_{e,0}\biggl(\frac{r}{r_0}\biggr)^{-\alpha}\biggl[1 + \biggl(\frac{r}{r_{e,0}}\biggr)^2\biggr]^{-3\beta_0/2} +\\
               &+ \rho_{e,1}\biggl[\biggl(\frac{r}{r_{e,1}}\biggr)^2\biggr]^{-3\beta_1/2}
\eeal
\ee
where: $\rho_{e,0}$ and $\rho_{e,1}$ are the normalization density constant; $r_{e,0}$ and $r_{e,1}$ are typical scale radii for each $\beta$-model component; and $r_0$ is the scale of the power-law truncated term of the first $\beta$-model appearing in Eq.~(\ref{eqn: beta model}). Note that the fit of the X-ray gas mass profiles is separately performed prior to the statistical inference of cluster halo (+DHOST) parameters (Section~\ref{subsec:stat}).
That is, $\{\rho_{e,0},\rho_{e,1},r_{e,0},r_{e,1},r_0,\alpha,\beta_1,\beta_2\}$ is not a free-parameter set but fixed to their respective best-fit values. 

Moreover, we must point out that when the double $\beta$-model and the truncated single one produce fits with the same statistical validity, for we do not have many data points or the data extend on a limited scale range, we opt to follow an Occam's razor approach and decide to work with the simplest model. Only in one case, MACSJ1720, we have found that the double $\beta$-model is highly preferable. 

\subsection{Brightest cluster galaxy}
\label{sec:BCG}

The availability of BCG data from the literature and the need for the most homogeneous data collection produces a cut on the full available CLASH sample.

Stellar mass estimates for the BCGs in CLASH clusters can be found in two works, \cite{2016ApJ...833..224C} and \cite{Burke:2015mda}, but with crucial differences for our purposes: in \cite{2016ApJ...833..224C} the authors performed a spectral energy distribution fitting using SDSS petrosian magnitudes and WISE band magnitudes as inputs and did not decompose the BCG contribution from the diffuse and low-brightness stellar component, the intracluster light (ICL), in cluster cores; in \cite{Burke:2015mda}, in contrast, such a decomposition was performed. These different procedures can unavoidably lead to significant differences between their BCG mass estimates up to $\sim 30\%$. Considering that we are to explicitly separate the BGC contribution from other member galaxies in our mass modelling, we have decided to consider BCG mass estimates from \cite{Burke:2015mda} as the most useful. As from Table~5 in \cite{Burke:2015mda}, we consider such masses evaluated at a distance of $r =50$~kpc from the center of the cluster.

Mass density modelling for the BCGs is then derived from photometric observations. Again, we have two main sources for such a data set, namely \cite{Tian:2020qjd} and \cite{2019A&A...622A..78D}. In \cite{Tian:2020qjd} the authors fit a single S\'{e}rsic profile to the surface brightness distribution of the BCG in the CLASH HST imaging data; in contrast, in \cite{2019A&A...622A..78D} both a S\'{e}rsic and a double S\'{e}rsic profile are used, and the best results (from a statistical point of view) are reported in their Table~2. For this reason, we have decided to focus on the results from \cite{2019A&A...622A..78D}. 

It must be stressed that we have checked and quantified the differences among our choices (i.e., mass estimates from \cite{Burke:2015mda} and photometry from \cite{2019A&A...622A..78D}) with also other combinations (i.e. using mass estimates from \cite{2016ApJ...833..224C} and/or photometry from \cite{Tian:2020qjd}). We have verified that if any difference is present, it is mainly due to the mass estimation and not to the photometric reconstruction.

We write the S\'{e}rsic profile as\footnote{Note that in \cite{2019A&A...622A..78D} the S\'{e}rsic profile is written in a different but totally equivalent way as:
\begin{equation}I(R) = I_{e} \exp \left[ -b_{n}\left( \frac{R}{R_{e}}\right)^{1/n}-1\right] \; ,
\end{equation}
with: $I_{e}$ the intensity at the effective radius $R_e$ (the half-light radius); $b_n$ a constant defined by $b_n = 2n-0.33$ \cite{Caon:1993wb}; and $n$ the S\'{e}rsic index.}
\begin{equation}\label{eq:sersic}
I(R) = I_{0} \exp \left[ -\left( \frac{R}{a_{s}}\right)^{1/n}\right] \; ,
\end{equation}
with: $I_{0}$ the central surface brightness; $a_{s}$ the S\'{e}rsic scale parameter; and $n$ the S\'{e}rsic shape parameter. The luminosity density can be obtained from Eq.~(\ref{eq:sersic}) using the approximation proposed by \cite{1997A&A...321..111P}:
\begin{equation}\label{eq:luminosity_density}
\ell(r) \equiv \ell_{1} \widetilde{\ell}(r/a_{s}) \; ,
\end{equation}
with:
\begin{equation}
\widetilde{\ell}(x) \simeq x^{-p} \, \exp(-x^{1/n}) \; ,
\end{equation}
\begin{equation}
\ell_{1} = \frac{L_\mathrm{tot}}{4 \pi \, n \, \Gamma[(3-p)n] a_{s}^{3}} \; ,
\end{equation}
where the function $p$ is defined in \citep{Neto:1999gx} as:
\begin{equation}
p \simeq 1.0 - 0.6097/n + 0.05463/n^2 \; ,
\end{equation}
and the total galaxy luminosity is
\begin{equation}\label{eq:ltot}
L_\mathrm{tot} \propto 10^{-0.4\,m} \; ,
\end{equation}
where $m$ is the apparent galaxy magnitude, and the proportionality takes into account the fact that other terms should add up in the exponential, such as the distance modulus or the magnitude calibration zero point. But we are mostly uninterested in their exact values, as we will fix them (or a combination of) directly when we need to adjust photometry to the total mass estimate.

In fact, once we have the photometric parameters $\{R_e,n,m\}$ from \cite{2019A&A...622A..78D}, we can model the BCG mass density profiles, in the most general case of two components, as
\be
\label{eqn: BCG}
\rho_\mathrm{BCG}(r) = \mathcal{A}\biggl[ l\left(\frac{r}{R_{e,int}}\right) +  l\biggl(\frac{r}{R_{e,ext}}\biggr)\biggr]
\ee
where the suffixes \textit{int} and \textit{ext} refer, in the notation of \cite{2019A&A...622A..78D}, to the fits of the central region and of the
BCG envelope, respectively, and the constant $\mathcal{A}$ contains the proportionality constant from Eq.~(\ref{eq:ltot}), the unknown mass-to-light ratio needed to convert luminosity densities into mass densities and all the factors required to obtain mass densities in units of $M_{\odot}$~kpc$^{-3}$. Such a constant is finally constrained by imposing that the total mass at $r=50$~kpc, derived from integrating Eq.~(\ref{eqn: BCG}), is set equal to the mass estimates from \cite{Burke:2015mda}.

\subsection{Galaxies}
\label{subsec:galaxies}

For the derivation of the galaxy density function excluding the BCG contribution, some preliminary steps are needed. Here we begin with the cold baryonic fraction $f_c(r)$ estimated in \cite{Chiu:2017nwm} based on  a sample of 91 clusters in the redshift range $0.25<z<1.25$. The cold baryonic fraction is defined as
\begin{equation}
f_{c} = \frac{M_{\ast}}{M_{\ast} + M_\mathrm{gas}}\, ,    
\end{equation}
where $M_{\ast}$ is the total stellar mass, which includes both the BCG contribution ($M_\mathrm{BCG}$) and non-BCG cluster galaxies ($M_\mathrm{gal}$) within the $r_{500}$ radius; and $M_\mathrm{gas}$ is the X-ray-emitting hot gas mass. The $f_c$ data are then fitted by the relation
\begin{equation}\label{eq:fc}
    f_c =\mathcal{A}_{f} \left( \frac{M_{500}}{M_\mathrm{piv}}\right)^{\mathcal{B}_f} \left( \frac{1+z}{1+z_\mathrm{piv}}\right)^{\mathcal{C}_f} \, ,
\end{equation}
where the pivot mass $M_\mathrm{piv}$ and redshift $z_\mathrm{piv}$ are the median values from the chosen cluster sample. The radial dependence of the obtained cold baryon fraction $ f_{c}(r)$ is shown in Fig.~11 of \cite{Chiu:2017nwm} as the black line.

It should be noted that the cold baryon fraction for our CLASH clusters cannot be determined without knowing the $r_{500}$ ($M_{500}$) value, which is a free parameter in our analysis, as stated in previous sections. We thus proceed in the following way: $i)$ we extract the radial trend $f_{c}(r)$ from Fig.~11 of \cite{Chiu:2017nwm} and we fit its logarithm with a polynomial of the third order in the variable $\log r$ (we ha tried many functional forms and this one provides the best agreement); $ii)$ we calculate the average cold baryon fraction $f_c$ for each cluster using Eq.~(\ref{eq:fc}) by providing $r_{500}$; $iii)$ we shift the relation from $i)$  to pass through the value found at step $ii)$ at $r=r_{500}$, and we use such a shifted relation as the best estimate for $f_{c}(r)$ of our CLASH clusters.

Then, starting from the definition of $f_{c}$ we find the mass profile of non-BCG galaxies as
\be
\label{eqn: gal}
M_\mathrm{gal}(<r) = \frac{f_c(r)}{1 - f_c(r)}M_\mathrm{gas}(<r) - M_\mathrm{BCG}(<r)
\ee
Finally, from \eqref{eqn: gal} we can derive non-BCG galaxy mass density as:
\begin{eqnarray}
\rho_\mathrm{gal}(r) &=& \frac{f_c(r)}{1 - f_c(r)}\rho_\mathrm{gas}(r) + \frac{M_\mathrm{gas}(<r)}{4\pi r^{2}} \left[ \frac{f_c(r)}{1 - f_c(r)}\right]' \nonumber \\ 
&-& \rho_\mathrm{BCG}(r) \, .
\end{eqnarray}

\subsection{Statistical analysis}
\label{subsec:stat}

To constrain the parameters of our model, we need to define the $\chi^2$ function for each probe we use and for each cluster. For the X-ray probe, the $\chi^2_\mathrm{gas}$ reads
\be
\label{eqn: chigas}
\chi^2_\mathrm{gas} = \sum_{i = 1}^\mathcal{N}\frac{\left[M^\mathrm{theo}_\mathrm{tot}(r_i,\boldsymbol{\theta}) - M^\mathrm{obs}_\mathrm{tot}(r_i)\right]^2}{\sigma_\mathrm{obs}^2(r_i)}\, ,
\ee
where $\mathcal{N}$ is the total number of data points we have at our disposal for each cluster \cite{Donahue:2014qda}; $r_i$ is the distance from the center of the cluster; $\sigma_\mathrm{obs}^2$ is the error associated to the total mass; and $\boldsymbol{\theta}$ is the vector of our free theoretical parameters.

As anticipated in the previous section, when working with GR, $\boldsymbol{\theta} = \{c_{500},r_{500} \}$, while when the DHOST model is considered, we have $\boldsymbol{\theta} = \{c_{500}, r_{500}, \alpha_H, \beta_1\}$.

For gravitational lensing, the $\chi_\mathrm{lens}^2$ function reads
\begin{equation}\label{eqn: chilens}
\chi^2_\mathrm{lens} = \boldsymbol{(\kappa^\mathrm{theo}(\theta)-\kappa^\mathrm{obs})}\cdot\mathbf{C}^{-1}\cdot\boldsymbol{(\kappa^\mathrm{theo}(\theta)-\kappa^\mathrm{obs})} \; ,
\end{equation}
where $\boldsymbol{\kappa^\mathrm{obs}}$ is the data vector containing the observed convergence measured in projected radial bins; $\boldsymbol{\kappa^\mathrm{theo}(\theta)}$ is the theoretical convergence calculated according to  Eq.~(\ref{eqn: kappamod}); and $\mathbf{C}$ is the total covariance matrix (see \cite{Umetsu:2015baa,Umetsu:2020wlf}).

The total $\chi^2$ function, defined as the sum of Eqs.~\eqref{eqn: chigas} and \eqref{eqn: chilens}, is then minimized using our own Monte Carlo Markov Chain (MCMC) code, whose convergence is checked using the method developed in \cite{Dunkley:2004sv}. Then, to provide a statistically meaningful and well characterized comparison between GR and DHOST using the same outputs of the MCMC method, we calculate the Bayesian evidence $\mathcal{E}$ for each model and for each cluster, using the nested sampling algorithm described in \cite{Mukherjee:2005wg}. We remind here that the Bayesian evidence $\mathcal{E}(D|\mathcal{M},\boldsymbol{\theta})$ is the degree of belief on data $D$ having a model $\mathcal{M}$ (GR and DHOST) and a set of parameters $\boldsymbol{\theta}$, and it is defined as
\be
\label{eqn: evidence}
\mathcal{E}(D|\mathcal{M},\boldsymbol{\theta}) = \int_{\Omega_\mathcal{M}}d\boldsymbol{\theta}\mathcal{L}(D|\boldsymbol{\theta},\mathcal{M})p(\boldsymbol{\theta}|\mathcal{M})
\ee
where: the integral is evaluated on the full parameter space $\Omega_\mathcal{M}$; $\mathcal{L}(D|\boldsymbol{\theta},\mt{M})$ is the likelihood function, directly connected to the total $\chi^2$ function by $\mathcal{L} \propto \exp \left(-\chi^2/2 \right)$; and $p(\boldsymbol{\theta}|\mathcal{M})$ is the prior probability of the model $\mathcal{M}$.

Since $\mathcal{E}$ depends on the choice of priors \cite{Nesseris:2012cq}, we have decided to use uninformative flat priors on all the involved free parameters, with the additional condition $\gamma_0>-1$, derived from from Eq.~(\ref{eq:GN_G}) in order to ensure that $G>0$.

Finally, we determine the Bayes Factor $\mathcal{B}$, defined as the ratio of evidence values of two models, $\mathcal{B}_{ij} = \mathcal{E}(\mathcal{M}_i)/\mathcal{E}(\mathcal{M}_j)$, where $\mathcal{M}_{j}$ is the reference model (in our case, GR). In general, the Bayes Factor is interpreted using an empirically calibrated scale, the Jeffrey's scale, which states that if $\ln\mathcal{B}_{ij} < 1$ the evidence in favour of model $\mathcal{M}_i$ is not significant with respect to the $\mathcal{M}_j$ one; if $1 < \ln\mathcal{B}_{ij} < 2.5$ the evidence is mild; if $2.5 < \ln\mathcal{B}_{ij} < 5$ the evidence is strong; if $\ln\mathcal{B}_{ij}> 5$ the evidence is decisive. Negative values of the Bayes factor are interpreted as evidence against the model $\mathcal{M}_i$ with respect to $\mathcal{M}_j$. 

\section{Results}
\label{sec: Results}

The first step in our analysis has been to compare our data with GR. This is important, first of all, because GR will be our reference model to which we can compare and estimate the statistical viability of the DHOST model. Next, it is because we can cross check our modelling and statistical analysis algorithm with results from the literature for the same sample. 

We have considered two different data combinations: lensing-only data ($\chi^2=\chi^2_\mathrm{lens}$); and lensing and X-ray hydrostatic data combined together ($\chi^2=\chi^2_\mathrm{lens}+\chi^2_\mathrm{gas}$). In \autoref{tab:results}, we report the resulting constraints on the parameters of the NFW density distribution, $c_{500}$ and $r_{500}$, for all 16 CLASH clusters ordered by crescent redshift $z$. 

Here we compare our results with those obtained in \cite{Umetsu:2015baa,Umetsu:2016cun,Chiu:2018gok}, reminding that in those works only lensing data were considered and thus a single NFW-component fitting was performed. Given the fact that our set of parameters  $\{c_{500}, r_{500}\}$ is not always chosen as the fitting set, with alternatives such as $\{c_{200}, r_{200}\}$ or $\{c_{200}, M_{200}\}$ often adopted, we have decided to convert both our parameter estimates and those from \cite{Umetsu:2016cun,Chiu:2018gok} into the characteristic NFW parameters, $\{\rho_{s}, r_s\}$, for a more direct comparison. Results are shown in Fig.~\ref{fig:NFW_comp_1}. We see that our estimates for $\rho_{s}$ (i.e. $c_{500}$) are in excellent agreement within the $1\sigma$ uncertainty with results from \cite{Umetsu:2016cun,Chiu:2018gok}. On the other hand, for what concerns $r_s$ (which depends on both $c_{500}$ and $r_{500}$), we see a larger deviation: our $r_s$ estimates are generally smaller than those from \cite{Umetsu:2016cun,Chiu:2018gok}; and only three clusters (MACSJ0329, MACSJ1149, and MACSJ0744) are not consistent within the $3\sigma$ confidence level with results from \cite{Chiu:2018gok} who performed a non-spherical, triaxial NFW fitting.

\begin{figure*}[htbp]
\centering
\includegraphics[width=8.5cm]{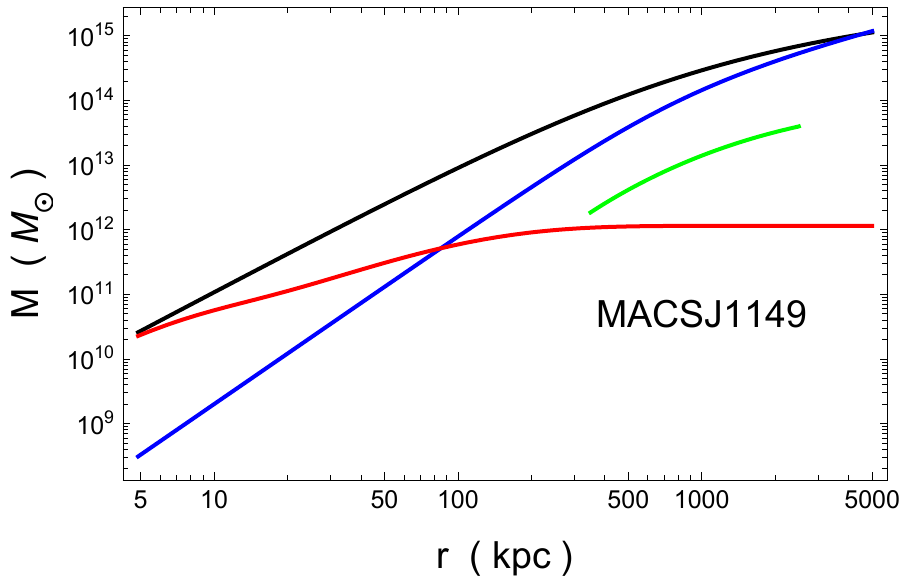}~~~
\includegraphics[width=8.5cm]{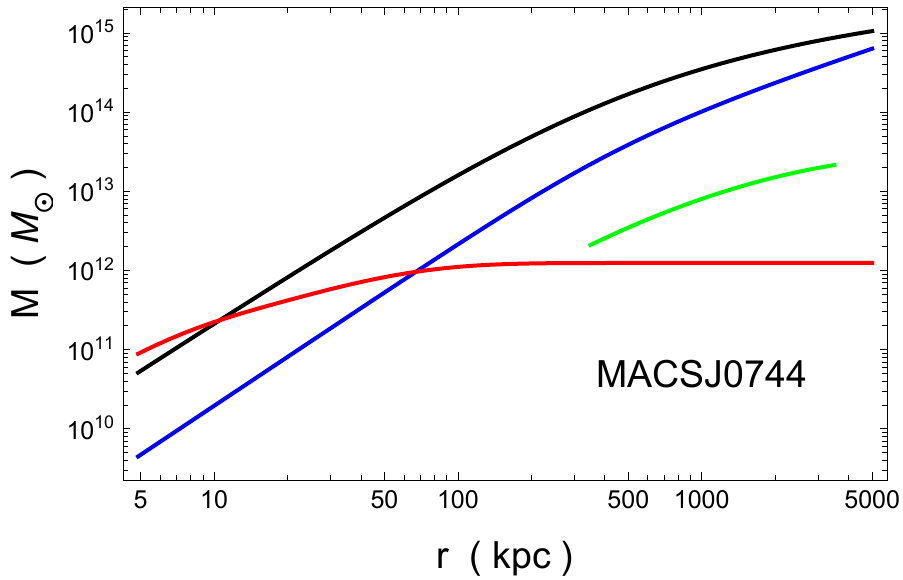}
\caption{Mass components considered in our modelling: NFW  dark matter - black; X-ray gas - blue; BCG - red; non-BCG galaxies - green.}\label{fig:mass_comp}
\end{figure*}

\begin{figure*}[htbp]
\centering
\includegraphics[width=17.5cm]{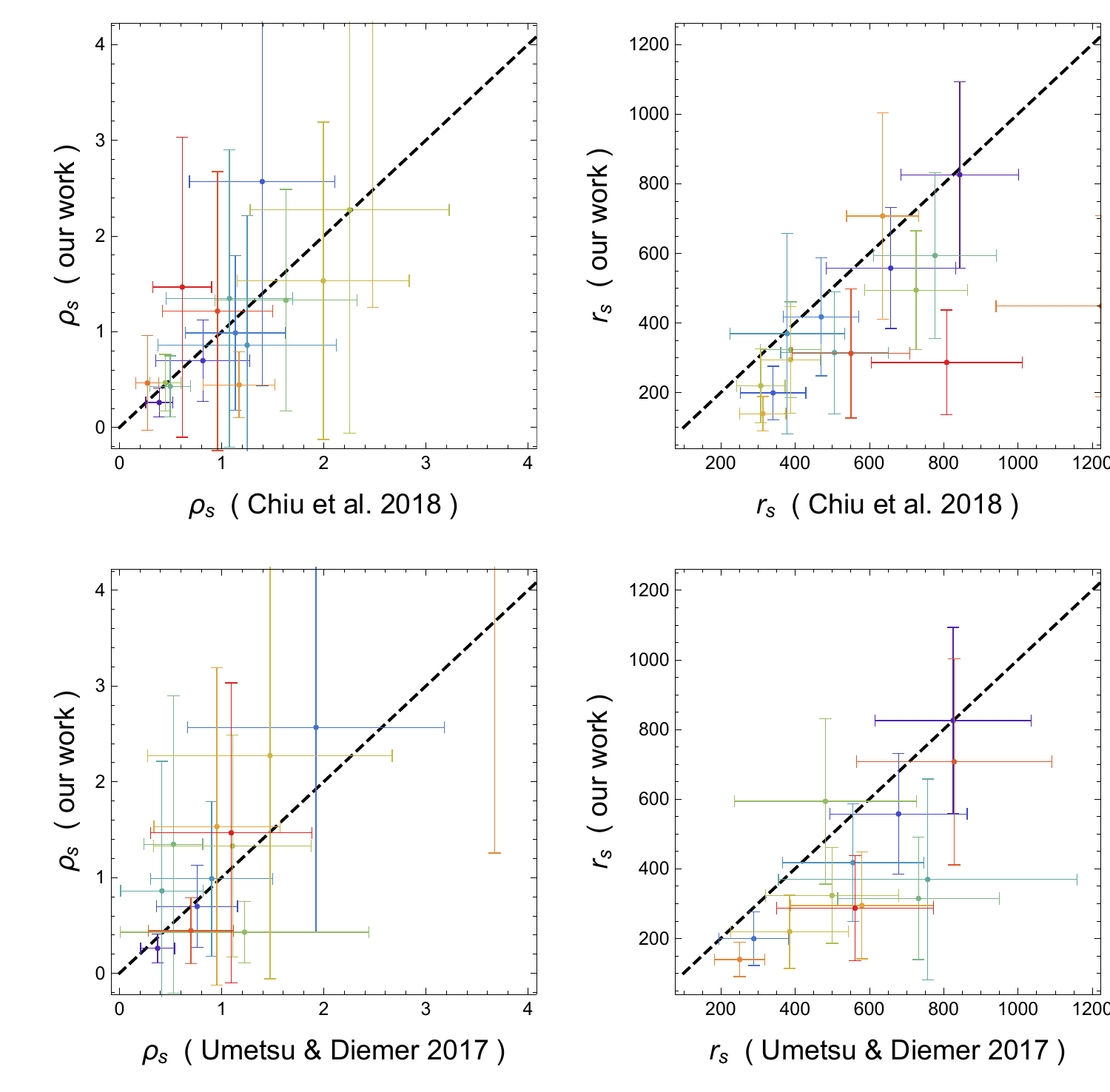}
\caption{Comparison of NFW parameters from our analysis with those derived in \cite{Umetsu:2016cun,Chiu:2018gok}.}\label{fig:NFW_comp_1}
\end{figure*}

One should stress that we carefully checked that this difference does not result from statistical bias in our MCMCM algorithm: the results are statistically consistent, convergence of the chains is reached, and the posteriors of the parameters do not have any pathological behaviour. Moreover, a visual inspection of Figs.~\ref{fig:fit_1} and \ref{fig:fit_2} clearly shows that there is no problem (within GR) to fit lensing data over the full range of scales from $10$~kpc to $\sim 5$~Mpc.  

Thus, any difference might be due to our choices in the mass modelling. One should note that in \cite{Umetsu:2015baa, Umetsu:2016cun} only the NFW component is fitted to the lensing signal, while we consider here a multi-component approach. That is, the NFW parameters derived from \cite{Umetsu:2016cun,Chiu:2018gok} represent the structural parameters of the "total" matter distribution, while those derived in this study describe the structure of the DM halo component.

Although the NFW component is the dominant contribution, we cannot neglect other mass components, as shown in Fig.~\ref{fig:mass_comp}. For instance, in the innermost part of the cluster (say, $r\lesssim 10-20$~kpc), the BCG component can be comparable and even dominant with respect to DM \cite{Okabe:2015qda,Sartoris:2020sdh}.  The only component that is directly related to $r_{500}$, apart from the DM component, is the non-BCG galaxy contribution. Here the $r_{500}$ radius essentially plays the role of a scaling parameter that can enhance or reduce the relative galaxy contribution. Non-BCG galaxies have a major influence in the radial range $50-200$~kpc, although being subdominant with respect to both DM and hot gas. 

Finally and importantly, intra-cluster gas physics too could induce a bias in the final estimation: being the density derived from X-ray observations, gas mass modelling can be affected by the non-local processes which we have described above. Thus, those clusters which exhibit disagreement between X-ray and lensing masses could lead to slightly different estimates for $r_{500}$. This seems to be the case, for example, of MACSJ1149 and MACSJ0744: according to Figs.~\ref{fig:fit_1} and \ref{fig:fit_2}, from a comparison of GR with combined data (dashed blue lines) and GR with lensing only (solid blue lines), we see how the above clusters exhibit a difference in the convergence profile in the inner regions ($r<100$~kpc).
It should be noted that MACSJ1149 is a highly disturbed lensing-selected CLASH cluster \cite{Postman:2011hg,Meneghetti2014}, while MACSJ0744 is the highest-redshift cluster in our sample. Both clusters were reported to have evidence of merger activity (see \cite{Postman:2011hg,Meneghetti2014,Sayers:2021zza}). 

\begin{figure*}[htbp]
\centering
\includegraphics[width=17.5cm]{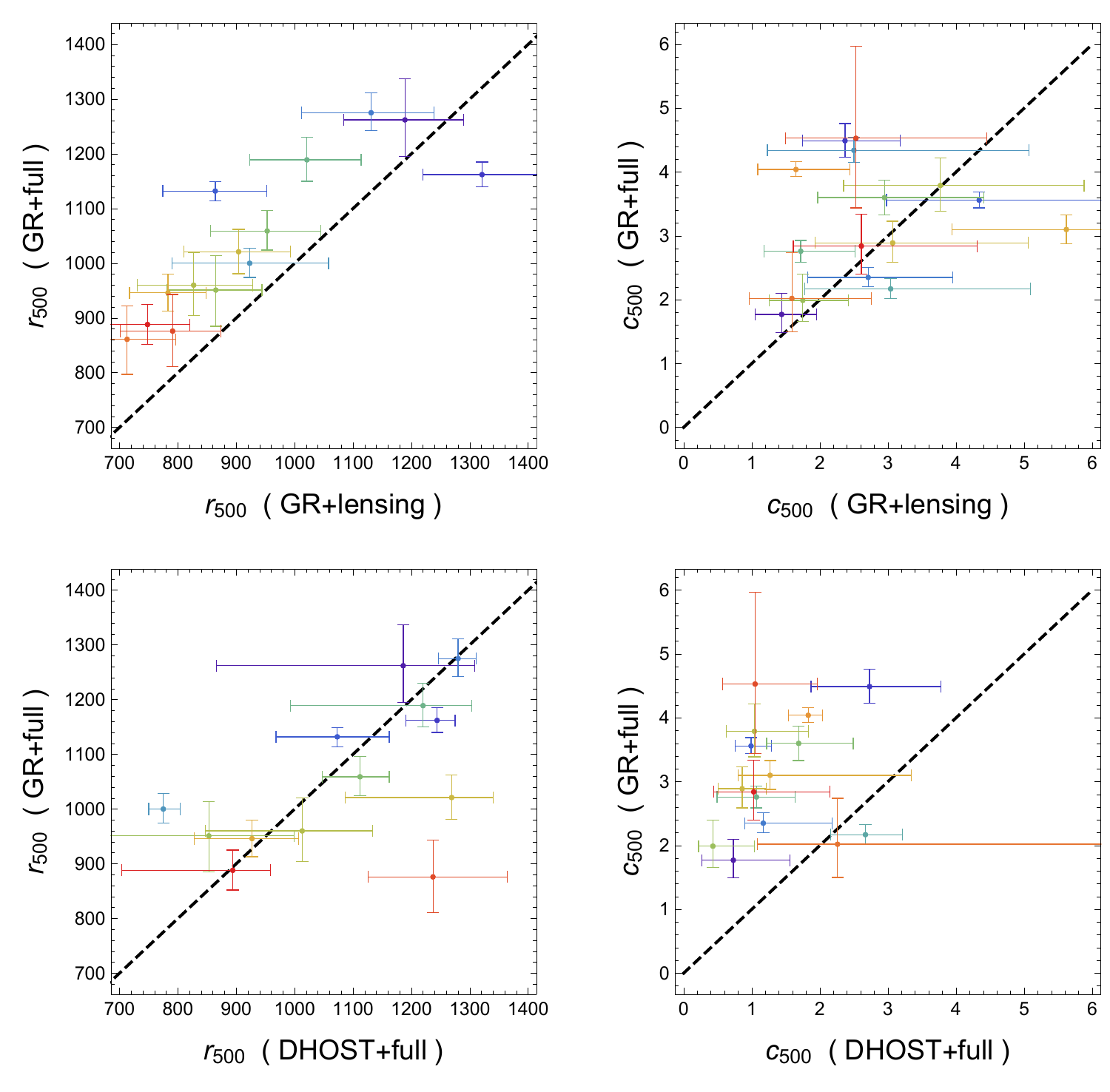}
\caption{Comparison of NFW parameters from GR analysis with those derived when a dark energy mimicking DHOST is considered.}\label{fig:DHOST_GR_1}
\end{figure*}

When it comes to a comparison between the lensing-only and the full data analysis (always within GR), we  see from the two top panels in Fig.~\ref{fig:DHOST_GR_1} how there is no statistically meaningful difference for what concerns the $c_{500}$ parameter, while there is a more striking bias in the $r_{500}$ estimates, although almost all clusters agree at the $<3\sigma$ level. However, this trend might be expected and be a signature of the above mentioned disagreement between lensing and X-ray based mass derivations.

Coming to our main goal in this work, which is to test the feasibility of the considered DHOST model in describing available data, we will consider two different scenarios:
\begin{enumerate}
    \item the DHOST model playing the role only of a DE alternative. This means that we will need explicitly a dark matter component in our analysis;
    \item the DHOST model as a possible description of both DE and DM at the galaxy clusters scale due to the partial breaking of the Vainshtein screening mechanism. In this case, the only mass components in our model will be hot gas, BCG and non-BCG galaxies.
\end{enumerate}

For what concerns the first scenario, we should first point out that an analysis by using lensing-only data (i.e., $\chi^2=\chi^2_\mathrm{lens}$) was not possible: possibly, observational errors are still too large to optimally constrain the model. By contrast, when using full data sets including X-ray information ($\chi^2=\chi^2_\mathrm{lens}+\chi^2_\mathrm{gas}$), we obtain more informative results. We should note that X-ray data have a much higher statistical precision than that of lensing data. On the other hand, the X-ray approach relies on the assumption of strict hydrostatic equilibrium, which may account for the apparent discrepancy between the two data sets.  In fact, in the context of $\Lambda$CDM, a mean hydrostatic mass bias of $b\equiv 1-M_\mathrm{X}/M_\mathrm{lens}=(12\pm 7)\%$  \cite{Donahue:2014qda} was found at $r=500$~kpc for the CLASH sample with respect to the weak-lensing mass estimates of \cite{Umetsu2014}. This is consistent with the typical level of hydrostatic mass bias ($5\%$--$20\%$) expected for cluster-scale haloes in $\Lambda$CDM cosmologies \cite{Nagai2007,Angelinelli2020}.

Results for the first scenario using full available data sets are listed in the second column group in Table~\ref{tab:results}. We report both median values (upper rows) and those corresponding to the minimum in the $\chi^2$ function (lower rows). From a Bayesian perspective, and considering the stochastic nature of MCMC sampling, the minimum $\chi^2$ values do not have a high statistical importance, but we have decided to show them as an indication for possible skewness in the posterior distribution and possible asymmetry among median and minimum estimates.

From the bottom panels of Fig.~(\ref{fig:DHOST_GR_1}), we see for the NFW parameters how we obtain $>3\sigma$ deviations for $r_{500}$ only in three cases, while we obtain systematically lower $c_{500}$ values compared to the GR case (in $14$ out of $16$ clusters). This suggests that the NFW dark matter haloes in the DHOST scenario are less concentrated and less massive in the inner regions of the clusters than what they are in the GR case, which is due to the partially broken screening mechanism effects.

\begin{figure*}[htbp]
\centering
\includegraphics[width=17.5cm]{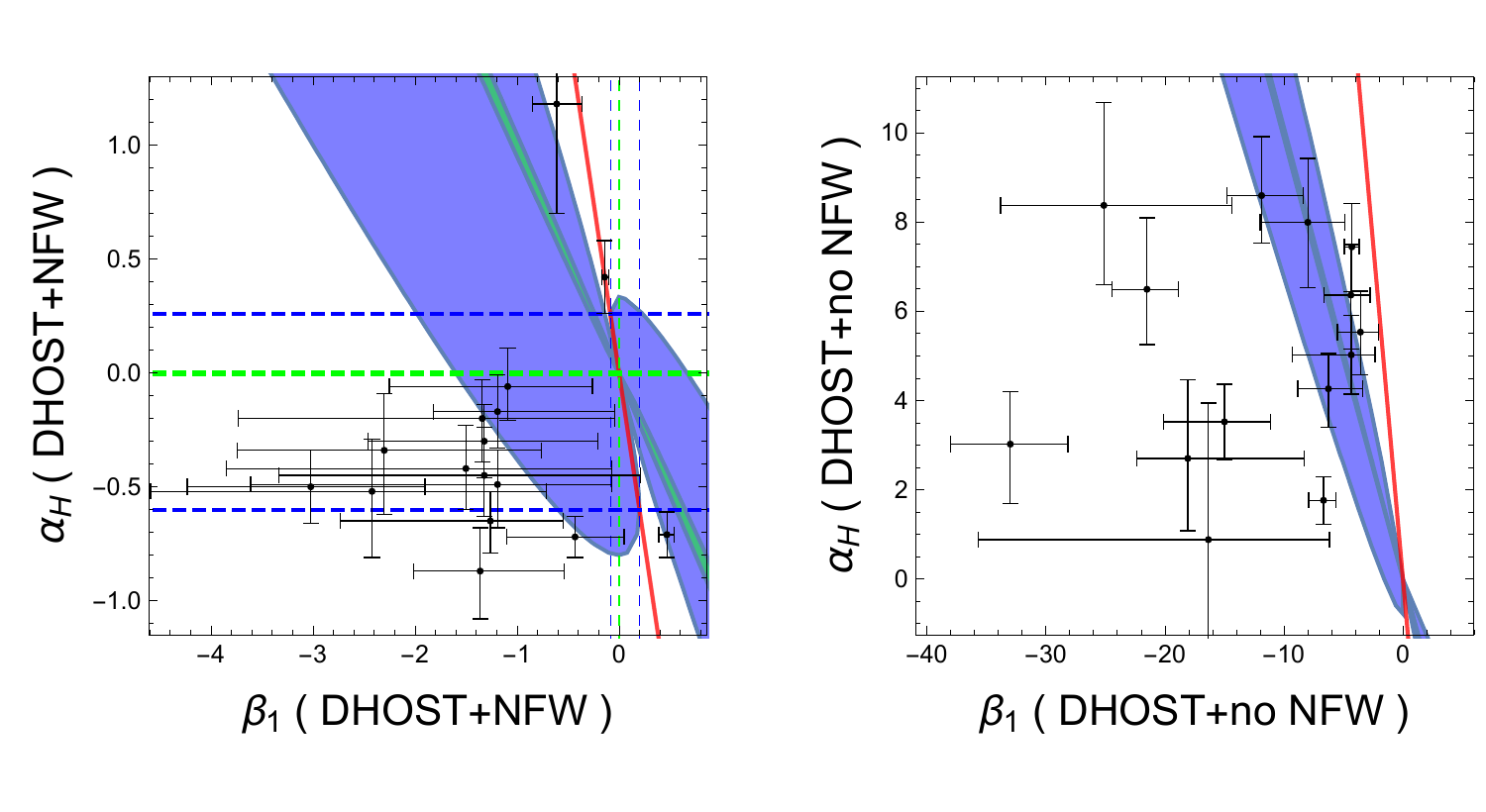}
\caption{Comparison of ETF parameters constraints from our dark energy mimicking DHOST model with results from \cite{Dima:2017pwp,Saltas:2019ius}.}\label{fig:EFT}
\end{figure*}

The DHOST model is characterized by the EFT parameters $\{\alpha_H,\beta_1\}$. In Fig.~(\ref{fig:EFT}) we compare our results with those derived from astrophysical constraints in \cite{Dima:2017pwp,Saltas:2019ius}. Blue regions show the constraints derived from stellar physics considerations, such as from hydrostatic equilibrium conditions or from the minimum mass of observed red dwarf stars; the red line is derived from limits on $\gamma_0$ from the Hulse-Taylor pulsar; the combination of these two constraints produces the range limited by the blue (vertical and horizontal) lines. When helioseismology is taken into account, we obtain the green regions, which combined with the prior on $\gamma_0$ leads to the region limited by the green (vertical and horizontal) lines. Much looser constraints can be derived from cosmological data \cite{Traykova:2019oyx,Hirano:2019nkz,Yamauchi:2021nxw} which are not shown in the figure.  

We can then see how the parameter $\alpha_H$ is quite well constrained and consistent with the previous limits, while $\beta_1$ is only marginally within the $1\sigma$ confidence limit for most of the clusters. What is even more intriguing is that our estimates for both parameters are only marginally consistent with the zero value (GR limit) in most of the cases at the $2\sigma$ level, at least.

Even so, we must stress two important points. The first point is visible from Figs.~(\ref{fig:fit_1}) and ~(\ref{fig:fit_2}): the DHOST model, together with an NFW dark matter component, fits quite well the data and is able to ``solve'' the discrepancy between X-ray and lensing which naturally appears in a GR context. This can be much clearly seen for the cases of, e.g., RXJ2248, MACSJ1115, MACS1720, and MACSJ0429, where the DHOST+NFW model (red line) is adjusting the GR+NFW with lensing only (solid blue line) much better than the GR+NFW with full data (dashed blue line) at smaller scales $(r<100$ kpc), while the situation reverses at larger scales $(r>1$ Mpc). Somehow, the terms arising from the broken screening mechanism are able to find a compromise among the different contributions leading to a sensibly better matching with data. This point is then furthermore confirmed by Table~\ref{tab:results2}, where the Bayes factors of the DHOST model with respect to GR are shown. We can see that in case of the DHOST+NFW model, the logarithm of the Bayes factor is always positive, ranging from inconclusive evidence in favour of DHOST (for values $\sim1$) to much more conclusive positive assessment (for values $>5$).

Last but not least, we examine the scenario where the DHOST model, together with its broken screening effects, is conjectured to behave as DM and DE at cluster scales. Thus, in this case, no NFW component is considered. Note that this model still includes the $r_{500}$ parameter that describes the non-BCG galactic component through the cold baryon fraction (Section~\ref{subsec:galaxies}).
Not having a full NFW profile to constrain in this case, we opt to apply a Gaussian prior on $r_{500}$ derived from the same corresponding GR scenario (i.e., with lensing-only data).

Here we are exploring a completely different approach from those in the literature, thus any comparison with constraints as shown above must be taken with care. 

First we point out that a good fit is not possible using full data sets ($\chi^2=\chi^2_\mathrm{lens}+\chi^2_\mathrm{gas}$),  whereas the modelling using lensing-only data ($\chi^2=\chi^2_\mathrm{lens}$) provides a reasonable fit, probably because of the much larger degrees of freedom given by the larger errors in lensing compared to the X-ray data. 

Now, both the EFT parameters are consistent with zero at the $3\sigma$ level, and they marginally agree with stellar constraints, excluding the Hulse-Taylor pulsar priors, as from the right panel of Fig.~(\ref{fig:DHOST_GR_1}). We exclude from this plot the two cases of MS2137 and MACSJ0329, which have some problematic issues in the profile fitting, as it can be seen from an inspection of the green lines and the curved profiles appearing at $r<500$~kpc. Other clusters as well show some problematic behaviours, as in MACSJ0647, for which we are not able to fit the data at $r>1$~Mpc.

In general, we find that such an approach is unable to describe the observed lensing profiles at large cluster-centric distances. In fact, the Bayes evidence ratios suggest that this model is significantly disfavoured with respect to GR.


\section{Conclusions}
\label{sec: Conclusions}

In this work, we have tested a particular DHOST model whose main characteristic is to exhibit a broken screening mechanism. Apart from playing the role as a dark energy alternative on cosmological scales, it may thus have influence on the internal dynamics of gravitational structures as large as clusters of galaxies. Taking this possibility to extremes, we also conjecture the possibility that such breaking could even lead to play the role of dark matter. That is, the DHOST might play the role of both dark energy and dark matter.

We focused our attention on a theory that has at most second-order terms described by \cite{Crisostomi:2017lbg}. As shown in Eqs.~(\ref{eqn: model}), this model is fully characterized, at the scales and in the limit of interest, by four parameters, namely,  $\{\Xi_1, \Xi_2, \Xi_3, \gamma_0\}$, which can be related to more general EFT parameters by Eqs.~(\ref{eqn: xi param}) \cite{Cardone:2020rmy}. This model represents a generalisation with respect to the one analysed in \cite{Salzano:2016udu,Salzano:2017qac}, including one more term in the metric potential equation.

We have tested the above DHOST model with a sample of 16 high-mass galaxy clusters targeted by the CLASH program \cite{Postman:2011hg} by combining two complementary probes: X-ray \cite{Donahue:2014qda} and strong-and-weak gravitational lensing \cite{Umetsu:2015baa} observations. In the X-ray approach, each cluster system is assumed to be in hydrostatic equilibrium. Moreover, we used a multi-component approach, modelling as many mass components as possible using observational results retrieved from the literature. We have been able to account for the hot gas, BCG, and non-BCG diffuse stellar contributions.

When the DHOST model is assumed to act as DE only, results show mild Bayesian evidence in favour of this model with respect to GR, for the majority of the clusters in our sample. What is more important is that in this scenario, we do not have any more discrepancy between X-ray hydrostatic and lensing mass measurements. Equivalently, we might say that apart from providing a better fit to the data than GR, the DHOST model is somehow able to reconcile such a discrepancy in a new theoretical approach. However, it should also be noted that the apparent discrepancy found in a GR context is likely to arise from the working hypothesis of hydrostatic equilibrium, which is not strictly satisfied in cluster haloes in the $\Lambda$CDM framework \citep{Nagai2007,Donahue:2014qda,Angelinelli2020}.  
In fact, about half of the sample clusters selected in this study are expected to be unrelaxed according to cosmological numerical simulations of \cite{Meneghetti2014}.

When the DHOST model is assumed to play the role of both DM and DE, through the partial breaking of the Vainshtein screening mechanism at cluster scales, the results show that this model is disfavoured with respect to GR. In fact, the Bayes Factors are negative and interpretable as evidence against such a DHOST model compared to GR.

{\renewcommand{\tabcolsep}{1.mm}
{\renewcommand{\arraystretch}{1.5}
\begin{table*}
\begin{minipage}{\textwidth}
\centering
\caption{CLASH clusters ordered by redshift. For each cluster, we provide $1\sigma$ constraints on each parameter in the top line, and the minimum values in the $\chi^2$ function in bottom one. Units: cluster radii are in kpc. For the cases of DHOST with no NFW component, we apply a Gaussian prior on $r_{500}$ derived from the corresponding GR cases. Lensing-selected CLASH clusters are indicated by stars.}\label{tab:results}
\resizebox*{\textwidth}{!}{
\begin{tabular}{c|cc|cc|cccc|ccc}
\hline
\hline
 & \multicolumn{4}{c|}{GR} & \multicolumn{4}{c|}{DHOST+NFW}  & \multicolumn{3}{c}{DHOST -  no NFW} \\
\hline
 & \multicolumn{2}{c|}{X-ray+lensing} 
 & \multicolumn{2}{c|}{lensing} 
 & \multicolumn{4}{c|}{X-ray+lensing}
 & \multicolumn{3}{c}{lensing} 
\\
 name 
 & $c_{500}$ & $r_{500}$ & $c_{500}$ & $r_{500}$  
 & $c_{500}$ & $r_{500}$ & $\alpha_{H}$  & $\beta_{1}$  
 & $r_{500}$ & $\alpha_{H}$ & $\beta_{1}$ \\
\hline
\hline
\multirow{2}{*}{A209} & 
$1.77^{+0.33}_{-0.28}$ & $1262^{+75}_{-67}$ & $1.44^{+0.51}_{-0.39}$ & $1189^{+100}_{-105}$ & 
$0.73^{+0.83}_{-0.46}$ & $1186^{+122}_{-320}$  & $-0.20^{+0.17}_{-0.19}$ & $-1.34^{+1.30}_{-2.39}$ &
$1189^{+107}_{-103}$ & $7.99^{+1.43}_{-1.47}$ & $-7.99^{+3.14}_{-4.00}$ \\
& $1.74$ & $1269$ & $1.33$ & $1219$ & $1.88$ & $1272$ & $-0.17$ & $0.38$ & $1189$ & $8.63$ & $-5.69$ \\
\hline
\multirow{2}{*}{A2261} & 
$4.49^{+0.27}_{-0.26}$ & $1162^{+23}_{-22}$  &  $2.37^{+0.81}_{-0.62}$ & $1321^{+100}_{-101}$ & 
$2.73^{+1.05}_{-0.86}$ & $1244^{+31}_{-53}$  & $-0.06^{+0.17}_{-0.15}$ & $-1.09^{+0.83}_{-1.16}$ &
$1320^{+99}_{-101}$ & $6.49^{+1.60}_{-1.24}$ & $-21.50^{+2.63}_{-2.92}$ \\
& $4.49$ & $1162$ & $2.22$ & $1348$ & $2.78$ & $1263$ & $-0.09$ & $-1.01$ & $1319$ & $7.10$ & $-18.43$ \\
\hline
\multirow{2}{*}{RXJ2129} & 
$3.56^{+0.13}_{-0.12}$ & $1132^{+17}_{-18}$  &  $4.34^{+1.87}_{-1.36}$ & $864^{+88}_{-90}$ &  
$0.99^{+0.30}_{-0.23}$ & $1073^{+89}_{-105}$ & $-0.50^{+0.16}_{-0.16}$ & $-3.02^{+1.12}_{-1.21}$ &
$867^{+91}_{-91}$ & $3.02^{+1.18}_{-1.33}$ & $-32.98^{+4.84}_{-5.00}$ \\
& $3.55$ & $1132$ & $4.03$ & $889$ & $1.31$ & $1169$ & $-0.48$ & $-1.81$ & $867$ & $3.74$ & $-25.52$ \\
\hline
\multirow{2}{*}{A611} & 
$2.35^{+0.16}_{-0.15}$ & $1275^{+36}_{-33}$  & $2.71^{+1.24}_{-0.89}$ & $1131^{+108}_{-119}$ &  
$1.17^{+1.01}_{-0.27}$ & $1280^{+31}_{-34}$  & $-0.17^{+0.16}_{-0.16}$ & $-1.19^{+1.15}_{-0.63}$ &
$1131^{+117}_{-119}$ & $8.37^{+2.31}_{-1.77}$ & $-25.12^{+10.75}_{-8.67}$ \\
& $2.35$ & $1276$ & $2.41$ & $1171$ & $0.97$ & $1272$ & $-0.20$ & $-1.60$ & $1130$ & $9.49$ & $-18.56$ \\
\hline
\multirow{2}{*}{MS2137} & 
$4.34^{+0.20}_{-0.19}$ & $1000.^{+28}_{-26}$ & $2.50^{+2.57}_{-1.27}$ & $923^{+135}_{-133}$ &  
$7.85^{+0.49}_{-0.52}$ & $775^{+29}_{-25}$ & $0.42^{+0.16}_{-0.16}$ & $-0.14^{+0.04}_{-0.03}$ &
$924^{+133}_{-131}$ & $-0.55^{+1.51}_{-1.41}$ & $-66.36^{+8.83}_{-6.78}$ \\
& $4.35$ & $999$ & $1.57$ & $1013$ & $7.73$ & $783$ & $0.11$ & $-0.05$ & $922$ & $0.99$ & $-55.93$ \\
\hline
\multirow{2}{*}{RXJ2248} & 
$2.17^{+0.16}_{-0.15}$ & $1547^{+47}_{-42}$ & $3.04^{+2.05}_{-1.26}$ & $956^{+119}_{-116}$ & 
$2.67^{+0.54}_{-0.51}$ & $1552^{+105}_{-96}$ & $-0.71^{+0.10}_{-0.10}$ & $0.47^{+0.07}_{-0.08}$ &
$959^{+116}_{-119}$ & $4.62^{+0.79}_{-0.87}$ & $-6.26^{+2.86}_{-2.58}$ \\
& $2.17$ & $1549$ & $2.53$ & $1005$ & $2.98$ & $1499$ & $-0.72$ & $0.45$ & $956$ & $4.60$ & $-6.23$ \\
\hline 
\multirow{2}{*}{MACSJ1115} & 
$2.76^{+0.17}_{-0.17}$ & $1189^{+41}_{-39}$ & $1.72^{+0.80}_{-0.54}$ & $1021^{+93}_{-98}$ &  
$1.07^{+0.57}_{-0.58}$ & $1220^{+83}_{-227}$ & $-0.49^{+0.15}_{-0.19}$ & $-1.19^{+1.12}_{-2.42}$ &
$1019^{+98}_{-99}$ & $2.70^{+1.77}_{-1.62}$ & $-18.07^{+9.77}_{-4.26}$ \\
& $2.76$ & $1188$ & $1.52$ & $1055$ & $2.58$ & $1243$ & $-0.48$ & $0.48$ & $1019$ & $4.30$ & $-8.67$\\
\hline
\multirow{2}{*}{MACSJ1720} & 
$3.60^{+0.27}_{-0.27}$ & $1059^{+37}_{-35}$ & $2.95^{+1.46}_{-0.98}$ & $953^{+92}_{-97}$ &  
$1.69^{+0.80}_{-0.47}$ & $1112^{+50}_{-65}$ & $-0.30^{+0.16}_{-0.16}$ & $-1.32^{+1.11}_{-1.14}$ &
$954^{+97}_{-97}$ & $8.59^{+1.32}_{-1.07}$ & $-11.87^{+3.47}_{-2.89}$ \\
& $3.59$ & $1060$ & $2.62$ & $984$ & $1.71$ & $1139$ & $-0.32$ & $-1.25$ & $950$ & $8.35$ & $-11.74$ \\
\hline
\multirow{2}{*}{MACSJ0416*} & 
$1.99^{+0.41}_{-0.33}$ & $951^{+63}_{-66}$ & $1.75^{+0.67}_{-0.49}$ & $865^{+79}_{-84}$ &  
$0.43^{+0.61}_{-0.21}$ & $853^{+146}_{-176}$ & $-0.34^{+0.25}_{-0.28}$ & $-2.30^{+1.54}_{-1.44}$ &
$869^{+77}_{-81}$ & $7.44^{+0.97}_{-1.00}$ & $-4.29^{+0.62}_{-0.63}$ \\
& $1.94$ & $960$ & $1.61$ & $888$ & $0.26$ & $761$ & $-0.49$ & $-3.04$ & $870$ & $7.31$ & $-4.20$ \\
\hline
\multirow{2}{*}{MACSJ0429} & 
$3.79^{+0.43}_{-0.40}$ & $960^{+60}_{-56}$ & $3.77^{+2.11}_{-1.42}$ & $827^{+101}_{-97}$ &  
$1.04^{+0.79}_{-0.41}$ & $1013^{+121}_{-166}$ & $-0.52^{+0.23}_{-0.29}$ & $-2.42^{+1.71}_{-2.17}$ &
$838^{+100}_{-100}$ & $0.88^{+3.06}_{-5.56}$ & $-16.34^{+10.18}_{-19.30}$ \\
& $3.77$ & $962$ & $3.24$ & $863$ & $3.92$ & $1012$ & $-0.47$ & $0.31$ & $833$ & $5.28$ & $-1.44$ \\
\hline
\multirow{2}{*}{MACSJ1206} & 
$2.89^{+0.34}_{-0.30}$ & $1201^{+41}_{-40}$ & $3.07^{+1.99}_{-1.14}$ & $904^{+89}_{-94}$ & 
$0.86^{+0.35}_{-0.35}$ & $1269^{+71}_{-182}$ & $-0.65^{+0.13}_{-0.14}$ & $-1.26^{+0.71}_{-1.47}$ &
$904^{+96}_{-96}$ & $3.52^{+0.84}_{-0.86}$ & $-14.99^{+3.88}_{-5.10}$ \\
& $2.87$ & $1204$ & $2.58$ & $942$ & $0.83$ & $1269$ & $-0.64$ & $-1.36$ & $904$ & $3.93$ & $-10.89$ \\
\hline
\multirow{2}{*}{MACSJ0329} & 
$3.10^{+0.23}_{-0.22}$ & $946^{+34}_{-33}$ & $5.62^{+2.16}_{-1.68}$ & $783^{+65}_{-66}$ &  
$1.27^{+2.07}_{-0.47}$ & $927^{+80}_{-99}$ & $-0.45^{+0.21}_{-0.18}$ & $-1.32^{+1.53}_{-2.01}$ &
$785^{+65}_{-64}$ & $-27.31^{+1.33}_{-1.75}$ & $-86.34^{+3.04}_{-5.44}$ \\
& $3.10$ & $946$ & $5.18$ & $798$ & $4.11$ & $900$ & $-0.27$ & $0.15$ & $784$ & $-29.47$ & $-93.20$ \\
\hline
\multirow{2}{*}{RXJ1347} & 
$4.04^{+0.12}_{-0.11}$ & $1508^{+23}_{-22}$ & $1.65^{+0.79}_{-0.56}$ & $1167^{+98}_{-107}$ & 
$1.83^{+0.21}_{-0.29}$ & $1681^{+29}_{-38}$ & $-0.72^{+0.09}_{-0.09}$ & $-0.43^{+0.48}_{-0.67}$ &
$1162^{+108}_{-107}$ & $5.53^{+0.92}_{-0.95}$ & $-3.57^{+1.54}_{-1.94}$ \\
& $4.04$ & $1509$ & $1.45$ & $1202$ & $1.99$ & $1696$ & $-0.72$ & $-0.21$ & $1169$ & $6.14$ & $-1.74$ \\
\hline
\multirow{2}{*}{MACSJ1149*} & 
$2.02^{+0.72}_{-0.52}$ & $861^{+61}_{-64}$ & $1.59^{+1.17}_{-0.63}$ & $713^{+83}_{-98}$ &  
$2.26^{+6.56}_{-1.18}$ & $410^{+126}_{-82}$ & $1.18^{+0.43}_{-0.48}$ & $-0.61^{+0.25}_{-0.24}$ &
$710.^{+99}_{-96}$ & $1.76^{+0.53}_{-0.53}$ & $-6.67^{+1.07}_{-1.24}$ \\
& $1.87$ & $879$ & $1.20$ & $753$ & $1.45$ & $716$ & $0.065$ & $-0.033$ & $711$ & $1.94$ & $-5.40$ \\
\hline
\multirow{2}{*}{MACSJ0647*} & 
$4.53^{+1.44}_{-1.09}$ & $876^{+67}_{-65}$ & $2.53^{+1.92}_{-1.03}$ & $791^{+83}_{-90}$ &  
$1.05^{+0.91}_{-0.48}$ & $1237^{+127}_{-111}$ & $-0.87^{+0.19}_{-0.21}$ & $-1.36^{+0.82}_{-0.65}$ &
$790^{+91}_{-90}$ & $6.36^{+1.13}_{-1.21}$ & $-4.37^{+1.62}_{-2.27}$ \\
& $4.26$ & $894$ & $2.04$ & $830$ & $0.75$ & $1319$ & $-0.98$ & $-1.42$ & $793$ & $6.55$ & $-4.59$ \\
\hline
\multirow{2}{*}{MACSJ0744} & 
$2.84^{+0.50}_{-0.44}$ & $888^{+37}_{-36}$ & $2.61^{+1.70}_{-1.00}$ & $748^{+72}_{-74}$ &  
$1.03^{+1.12}_{-0.59}$ & $894^{+65}_{-190}$ & $-0.42^{+0.19}_{-0.18}$ & $-1.50^{+1.43}_{-2.35}$ &
$748^{+74}_{-74}$ & $5.02^{+0.89}_{-0.88}$ & $-4.34^{+2.00}_{-4.95}$ \\
& $2.82$ & $891$ & $2.09$ & $777$ & $2.95$ & $919$ & $-0.47$ & $0.32$ & $747$ & $4.55$ & $-10.42$ \\
\hline
\hline
\end{tabular}}
\end{minipage}
\end{table*}}}

{\renewcommand{\tabcolsep}{1.mm}
{\renewcommand{\arraystretch}{1.5}
\begin{table*}
\begin{minipage}{\textwidth}
\centering
\caption{CLASH clusters ordered by redshift. Units: cluster radii are in kpc. For the cases of DHOST with no NFW component, we apply a Gaussian prior on $r_{500}$ derived from the corresponding GR cases. Lensing-selected CLASH clusters are indicated by stars.}\label{tab:results2}
\resizebox*{0.85\textwidth}{!}{
\begin{tabular}{c|cc|ccc|ccc}
\hline
\hline
 & \multicolumn{2}{c|}{GR} & \multicolumn{3}{c|}{DHOST+NFW}  & \multicolumn{3}{c}{DHOST - no NFW} \\
\hline
 & X-ray+lensing & lensing & \multicolumn{3}{c|}{X-ray+lensing} & \multicolumn{3}{c}{lensing} \\
name & $\chi^{2}$ & $\chi^{2}$ & $\chi^{2}$ & $\mathcal{B}^\mathrm{DHOST}_\mathrm{GR}$ & $\ln \mathcal{B}^\mathrm{DHOST}_\mathrm{GR}$ & 
$\chi^{2}$ & $\mathcal{B}^\mathrm{DHOST}_\mathrm{GR}$ & $\ln \mathcal{B}^\mathrm{DHOST}_\mathrm{GR}$ \\
\hline
A209     & $10.69$ &  $8.64$ &  $8.50$ & $1.56^{+0.05}_{-0.04}$ & $0.44^{+0.03}_{-0.03}$ & $11.02$ & $0.222^{+0.007}_{-0.005}$ & $-1.51^{+0.03}_{-0.02}$ \\
A2261     & $11.33$ &  $5.43$ &  $9.64$ & $1.08^{+0.03}_{-0.03}$ & $0.08^{+0.03}_{-0.02}$ & $10.00$ & $0.072^{+0.002}_{-0.002}$ & $-2.63^{+0.02}_{-0.02}$ \\
RXJ2129   & $20.90$ &  $7.28$ & $7.94$ &  $254^{+8}_{-8}$ & $5.54^{+0.03}_{-0.03}$ & $8.29$ &  $0.43^{+0.01}_{-0.01}$ & $-0.86^{+0.02}_{-0.02}$ \\
A611      &  $5.84$ &  $3.41$ &  $4.18$ &  $1.07^{+0.03}_{-0.02}$ & $0.06^{+0.03}_{-0.02}$ &  $4.04$ &  $0.53^{+0.01}_{-0.01}$ & $-0.63^{+0.02}_{-0.02}$ \\
MS2137    & $34.38$ &  $7.74$ & $13.51$ & $\left(1.77^{+0.06}_{-0.05}\right) \cdot 10^{4}$ & $9.78^{+0.03}_{-0.03}$ &  $5.18$ & $2.87^{+0.08}_{-0.06}$ & $1.05^{+0.03}_{-0.02}$ \\
RXJ2248   & $58.93$ &  $4.54$ &  $6.99$ & $\left(7.62^{+0.20}_{-0.17}\right) \cdot 10^{10}$ & $25.06^{+0.03}_{-0.02}$ &  $5.91$ & $0.363^{+0.008}_{-0.010}$ & $-1.01^{+0.02}_{-0.03}$\\
MACSJ1115 & $19.67$ &  $7.12$ &  $7.21$ & $235^{+5}_{-7}$ & $5.46^{+0.02}_{-0.03}$ & $15.50$ &  $\left(1.15^{+0.02}_{-0.03}\right) \cdot 10^{-2}$ & $-4.47^{+0.02}_{-0.03}$\\
MACSJ1720 & $10.44$ &  $5.00$ &  $5.93$ & $6.20^{+0.19}_{-0.17}$ & $1.82^{+0.03}_{-0.03}$ & $4.99$ & $0.74^{+0.02}_{-0.02}$ & $-0.30^{+0.02}_{-0.03}$ \\
MACSJ0416* & $14.86$ &  $8.47$ &  $9.38$ & $32.6^{+1.0}_{-1.0}$ & $3.48^{+0.03}_{-0.03}$ & $12.04$ & $0.127^{+0.004}_{-0.003}$ & $-2.06^{+0.03}_{-0.02}$ \\
MACSJ0429 & $11.76$ &  $5.98$ &  $6.50$ & $8.90^{+0.27}_{-0.27}$ & $2.19^{+0.03}_{-0.03}$ & $16.10$ & $\left(1.03^{+0.03}_{-0.03}\right) \cdot 10^{-2}$ & $-4.58^{+0.02}_{-0.03}$ \\
MACSJ1206 & $31.93$ &  $8.07$ &  $7.95$ & $\left(67.4^{+2.4}_{-1.5}\right)\cdot10^3$ & $11.12^{+0.03}_{-0.02}$ &  $7.14$ & $1.21^{+0.04}_{-0.02}$ & $0.19^{+0.03}_{-0.02}$ \\
MACSJ0329 & $20.76$ & $10.53$ &  $11.17$ & $72.8^{+2.6}_{-2.0}$ & $4.29^{+0.04}_{-0.03}$ & $35.82$ & $\left(2.18^{+0.06}_{-0.06}\right) \cdot 10^{-6}$ & $-13.04^{+0.02}_{-0.03}$ \\
RXJ1347   & $70.82$ &  $3.67$ &  $6.40$ & $\left(1.27^{+0.03}_{-0.02}\right) \cdot 10^{14}$ & $32.48^{+0.03}_{-0.02}$ &  $5.31$ & $0.50^{+0.01}_{-0.01}$ & $-0.69^{+0.03}_{-0.03}$  \\
MACSJ1149* & $31.20$ &  $8.06$ &  $8.81$ & $\left(16.98^{+0.61}_{-0.34}\right) \cdot 10^3$ & $9.74^{+0.04}_{-0.02}$ &  $8.93$ & $0.52^{+0.01}_{-0.01}$ & $-0.66^{+0.03}_{-0.03}$\\
MACSJ0647* & $24.92$ &  $6.16$ &  $5.81$ & $\left(3.61^{+0.12}_{-0.09}\right) \cdot 10^3$ & $8.19^{+0.03}_{-0.03}$ & $11.78$ & $\left(4.56^{+0.09}_{-0.10}\right) \cdot 10^{-2}$ & $-3.09^{+0.02}_{-0.02}$ \\
MACSJ0744 & $17.13$ &  $9.93$ &  $10.01$ & $190^{+4}_{-4}$ & $5.25^{+0.02}_{-0.02}$ & $10.90$ & $0.46^{+0.01}_{-0.01}$ & $-0.77^{+0.03}_{-0.03}$\\
\hline
\hline
\end{tabular}}
\end{minipage}
\end{table*}}}

\clearpage
\newpage

\begin{figure*}[htbp]
\centering
\includegraphics[width=8.5cm]{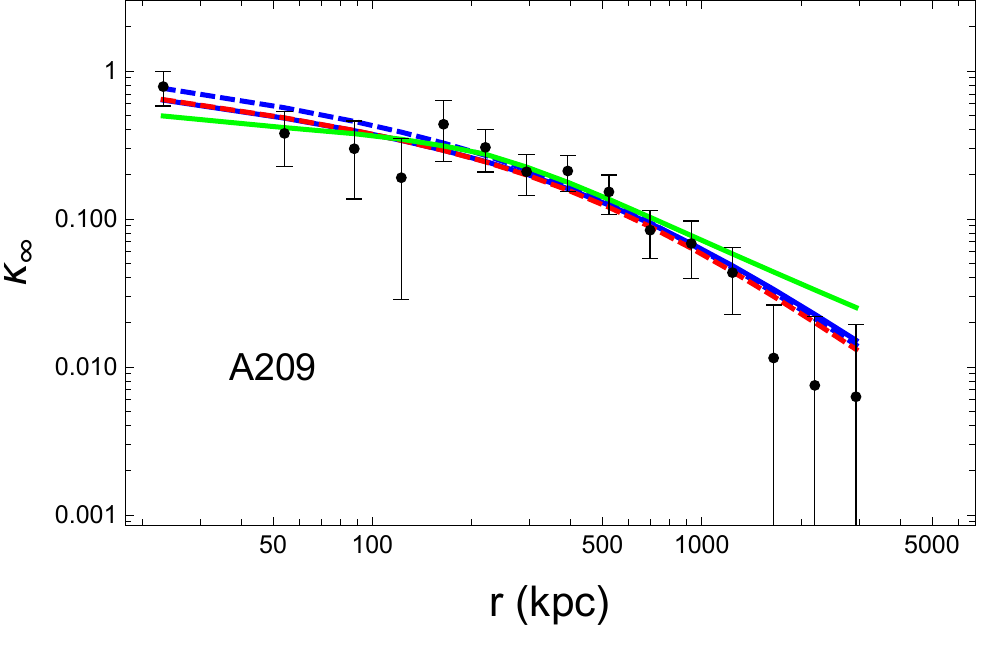}~~~
\includegraphics[width=8.5cm]{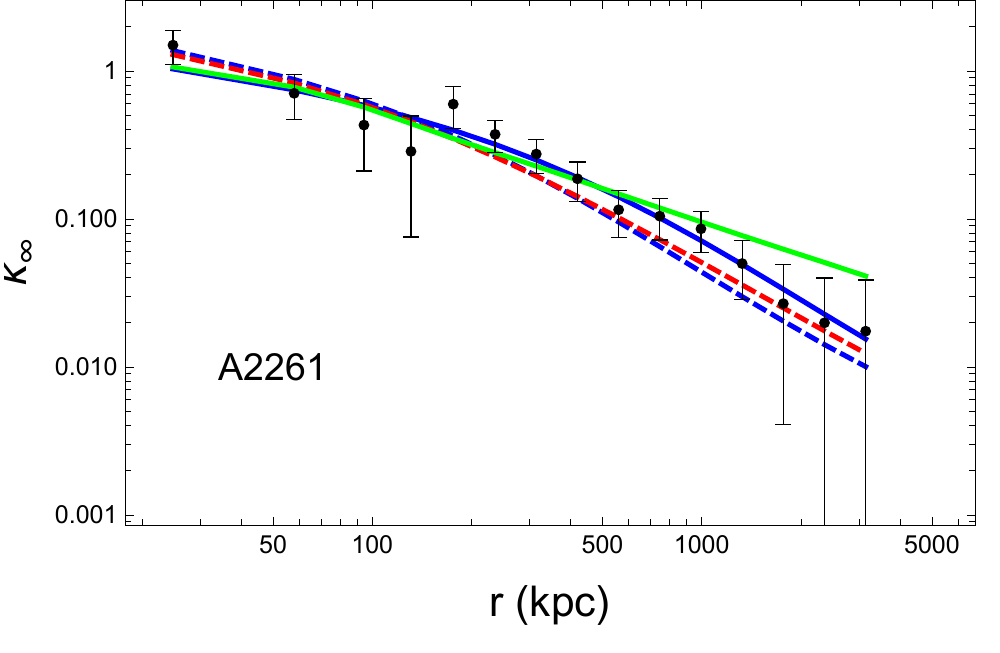}\\
\includegraphics[width=8.5cm]{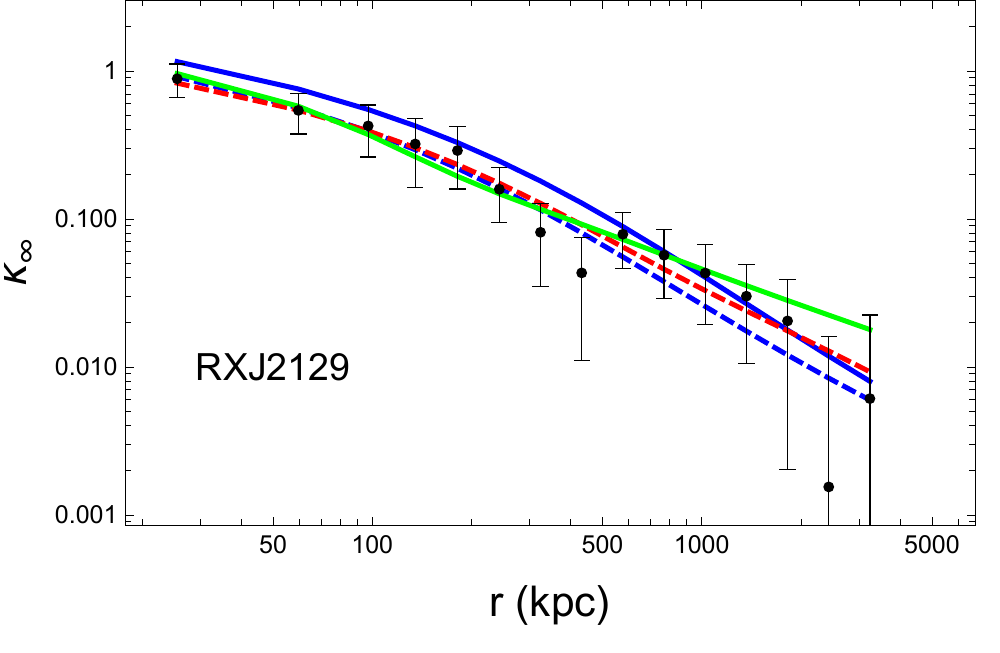}~~~
\includegraphics[width=8.5cm]{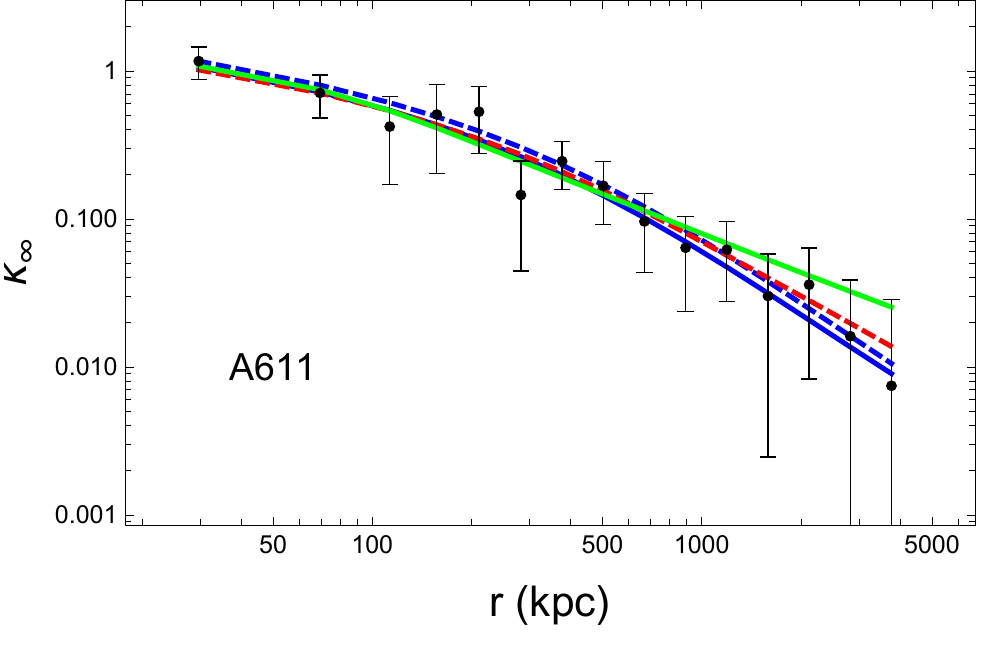}\\
\includegraphics[width=8.5cm]{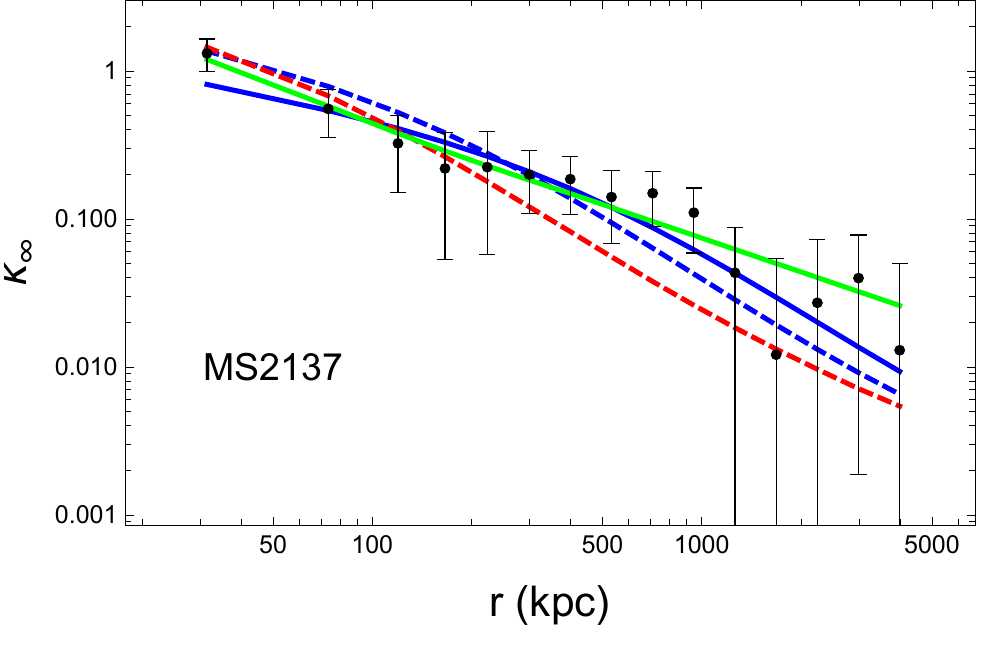}~~~
\includegraphics[width=8.5cm]{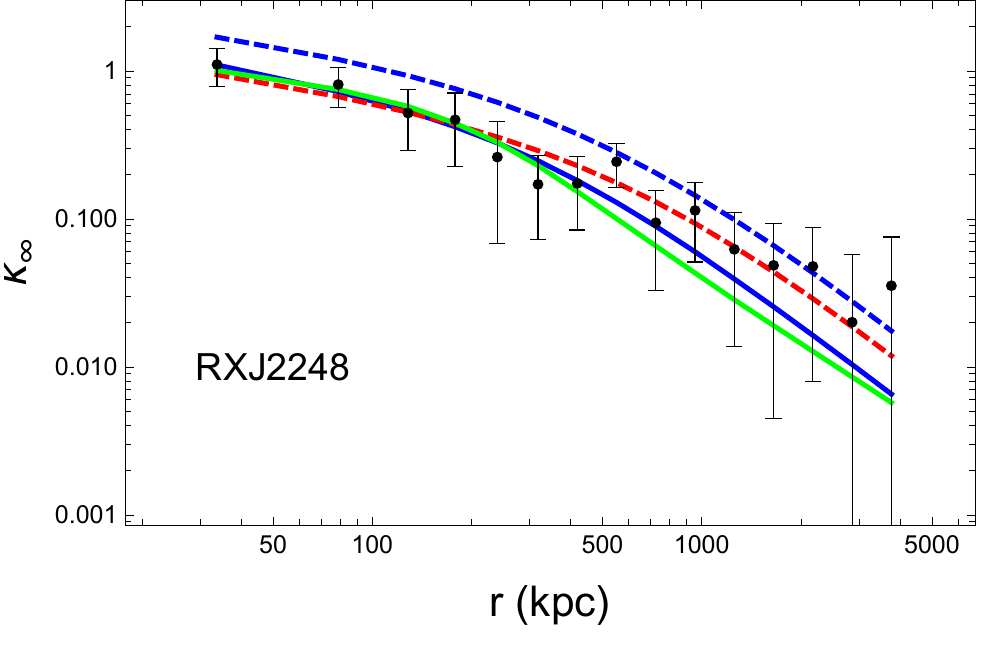}\\
\includegraphics[width=8.5cm]{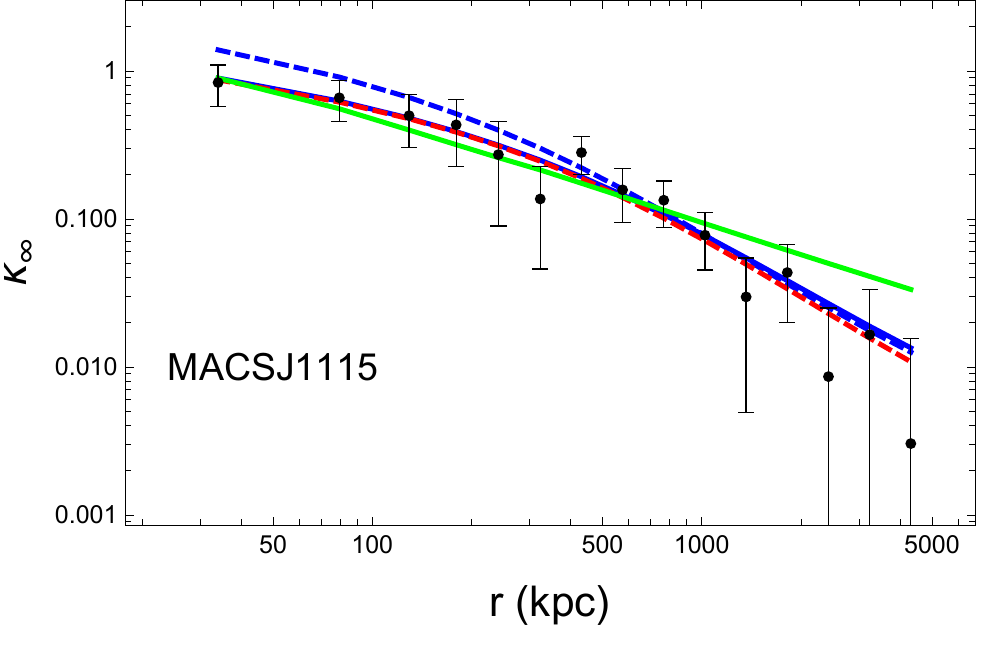}~~~
\includegraphics[width=8.5cm]{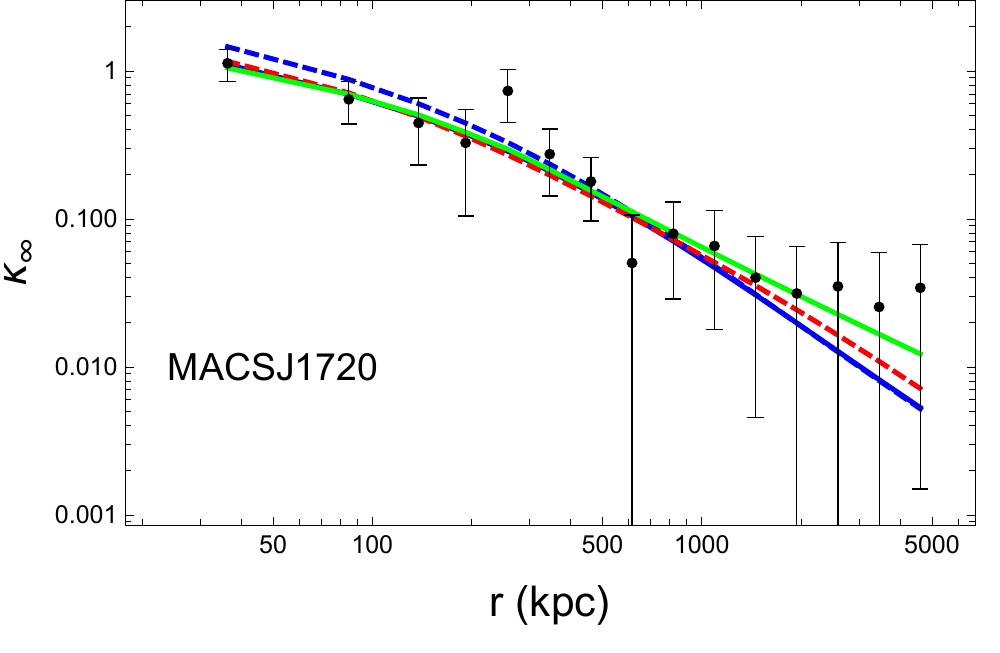}\\
\caption{Convergence profiles reconstructed from gravitational lensing. Color code: black points - observational data from \cite{Umetsu:2015baa}; blue dashed - GR (all), blue solid - GR (lensing); dashed red - DHOST + NFW (all); green - DHOST (no NFW, lensing).}\label{fig:fit_1}
\end{figure*}

\begin{figure*}[htbp]
\centering
\includegraphics[width=8.5cm]{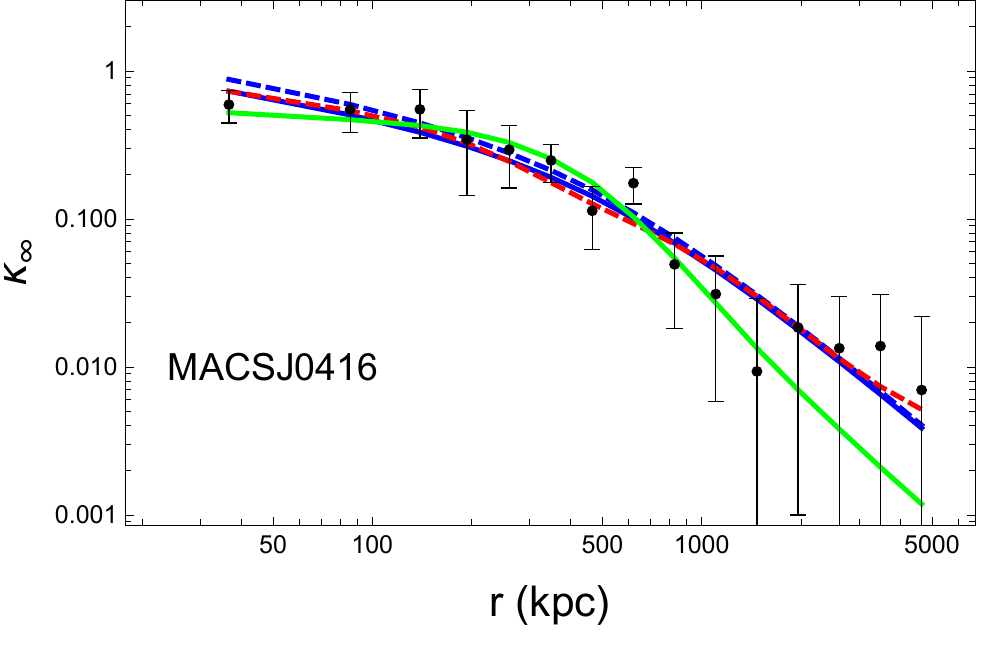}~~~
\includegraphics[width=8.5cm]{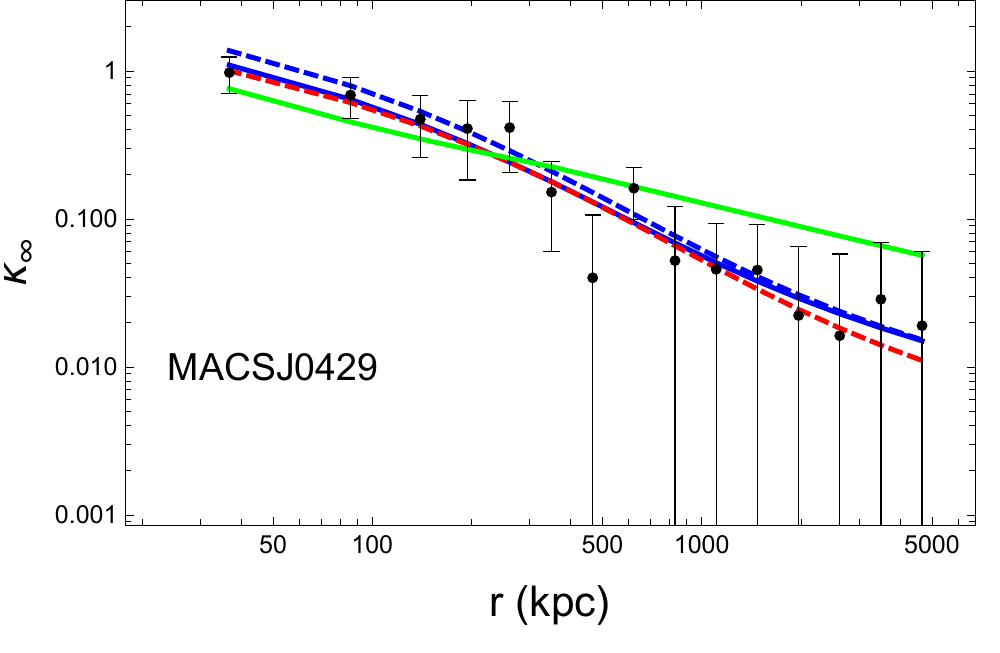}\\
\includegraphics[width=8.5cm]{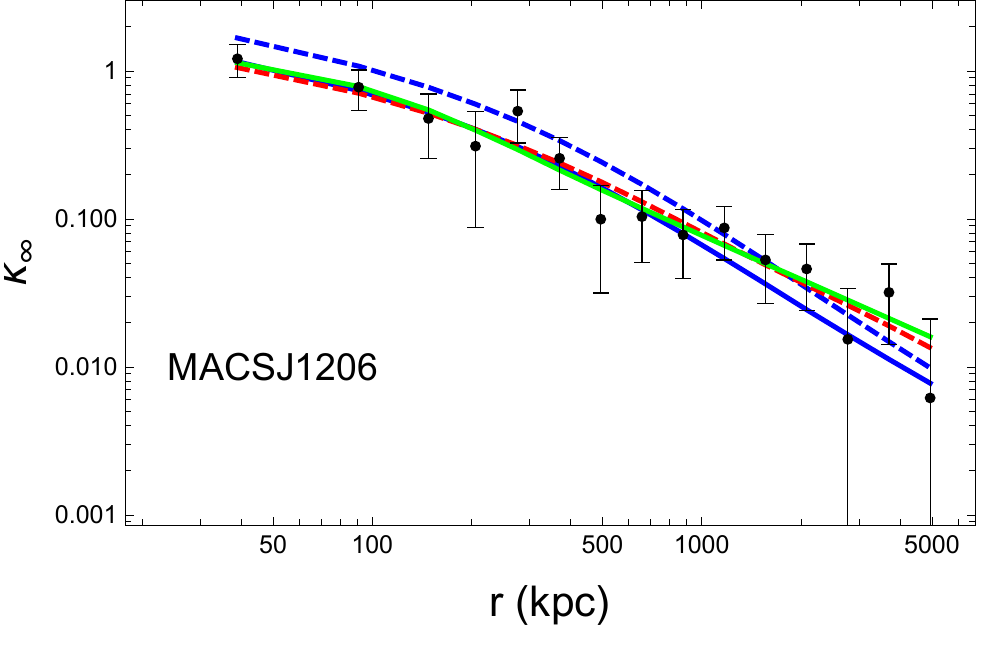}~~~
\includegraphics[width=8.5cm]{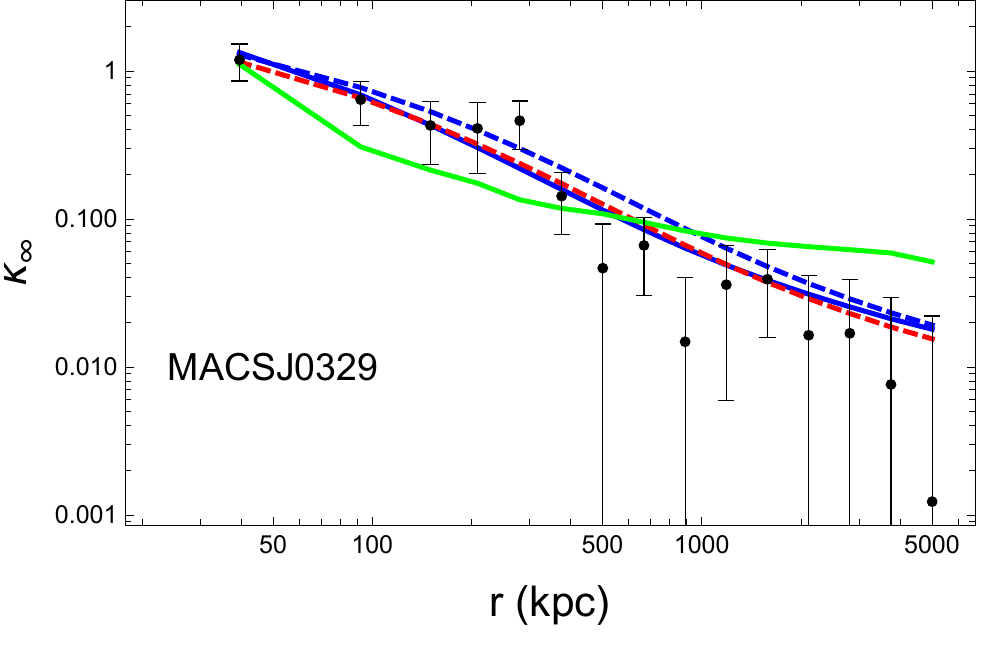}\\
\includegraphics[width=8.5cm]{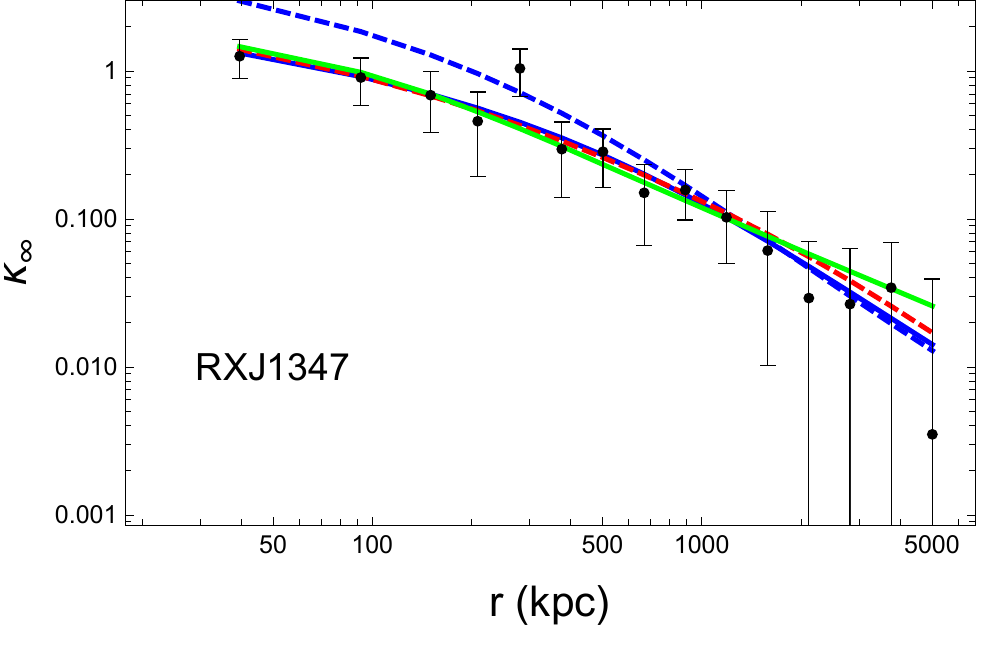}~~~
\includegraphics[width=8.5cm]{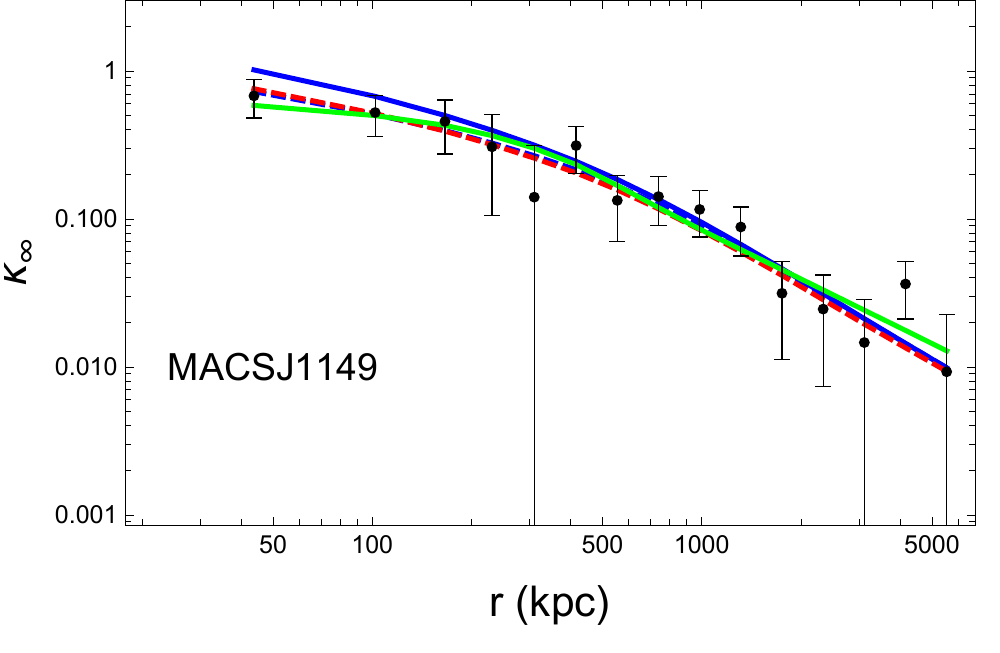}\\
\includegraphics[width=8.5cm]{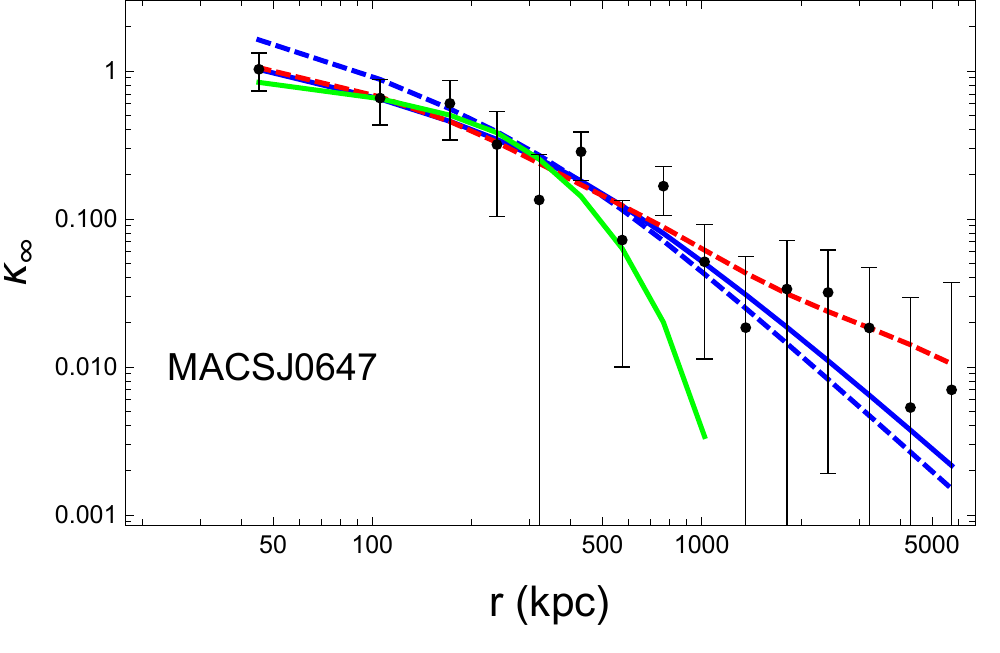}~~~
\includegraphics[width=8.5cm]{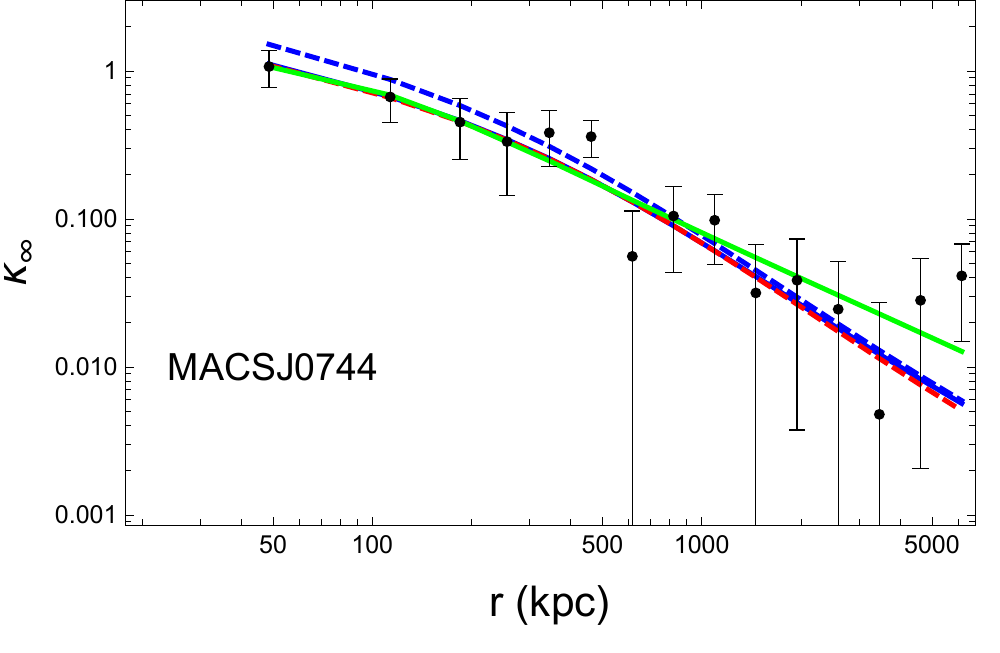}\\
\caption{Convergence profiles reconstructed from gravitational lensing. Color code: black points - observational data from \cite{Umetsu:2015baa}; blue solid - GR (all), blue dashed - GR (lensing); red - DHOST + NFW (all); green - DHOST (no NFW, lensing).}\label{fig:fit_2}
\end{figure*}

\bibliographystyle{apsrev4-1}
\bibliography{biblio}
\end{document}